\documentclass[final,3p,sort,compress,times]{elsarticle}

\usepackage{multirow,setspace,times,amssymb,amsmath,graphicx,color,rotating,subfigure,url}

\graphicspath{{figures/}}

\journal{Physica A}

\begin{document}

\begin{frontmatter}


\title{Wax and wane of the cross-sectional momentum and contrarian effects: \\Evidence from the Chinese stock markets}

\author[BS,RCE]{Huai-Long Shi}
\author[BS,RCE,SS]{Wei-Xing Zhou\corref{cor1}}
\cortext[cor1]{Corresponding author. Address: 130 Meilong Road, P.O. Box 114, School of Business, East China University of Science and Technology, Shanghai 200237, China, Phone: +86-21-64253634.}
\ead{wxzhou@ecust.edu.cn}%

\address[BS]{Department of Finance, School of Business, East China University of Science and Technology, Shanghai 200237, China}
\address[RCE]{Research Center for Econophysics, East China University of Science and Technology, Shanghai 200237, China}
\address[SS]{Department of Mathematics, School of Science, East China University of Science and Technology, Shanghai 200237, China}

\begin{abstract}
  This paper investigates the time-varying risk-premium relation of the Chinese stock markets within the framework of cross-sectional momentum and contrarian effects by adopting the Capital Asset Pricing Model and the French-Fama three factor model. The evolving arbitrage opportunities are also studied by quantifying the performance of time-varying cross-sectional momentum and contrarian effects in the Chinese stock markets. The relation between the contrarian profitability and market condition factors that could characterize the investment context is also investigated. The results reveal that the risk-premium relation varies over time, and the arbitrage opportunities based on the contrarian portfolios wax and wane over time. The performance of contrarian portfolios are highly dependent on several market conditions. The periods with upward trend of market state, higher market volatility and liquidity, lower macroeconomics uncertainty are related to higher contrarian profitability. These findings are consistent with the Adaptive Markets Hypothesis and have practical implications for market participants.
\end{abstract}


\end{frontmatter}


\section{Introduction}
\label{S1:Intro}

The researchers adhering to classic financial theories that advocate unbounded rationality attempt to add new risk factors to asset pricing model so as to explain the abnormal
returns associated with market anomalies. Other researchers who stick to the bounded rationality of traders try to explain the market anomalies through constructing behavioural models capable of capturing the behavioural biases of traders. Nevertheless, both streams of researchers fail to provide the satisfactory solution to explaining these anomalies. Although abnormal returns related to most of the anomalies disappear in more complete pricing models with more risk factors, some anomalies still ``survive'' \cite{Fama-French-1996-JF}.
On the other hand, though anomaly puzzles such as short-term momentum effect could be solved as the behavioural biases of traders within the framework of behavioural finance, not a unified model could cover more anomalies. As documented in Ref.~ \cite{Lo-2004-JPM}, the prospect theory depicts the loss aversion, but fails to portray the overconfidence \cite{Kahneman-Tversky-1979-Em}.

A more comprehensive theory needs be developed to resolve the conflict between the Efficient Markets Hypothesis (EMH) \cite{Fama-1970-JF,Fama-1991-JF} and behavioural finance \cite{Barberis-Shleifer-Vishny-1998-JFE,Hong-Stein-1999-JF,Daniel-Hirshleifer-Subrahmanyam-1998-JF}. Based on evolutionary psychology and evolutionary theory, the Adaptive Markets Hypothesis (AMH) came out \cite{Lo-2004-JPM,Lo-2005-JIC}. As with the EMH, traders behave based on their own interest under the AMH. It is a process of heuristics in the adaptive markets where traders are allowed to make mistakes and gradually adapt themselves to changing market status by correcting their own mistakes. By comparison, traders are such perfectly rational that market has maintained in a state of equilibrium. In addition, under the AMH, different trader groups have the competition for scarce profitability opportunities, thus leading to the ``survival of richest''. What ultimately plays a decisive role is the natural selection. The evolving market participants mainly contribute to the time-varying or context-dependent market efficiency. As stated in the AMH, efficient markets can be regarded as an ideal and equilibrium state with fixed proportion of traders types and fixed market conditions, while the behavioural biases of traders characterize the adaptive process for the transition of market status. Accordingly, the EMH can co-exist with behavioural finance within this new paradigm.

The essential inference of the AMH is that market efficiency is not an all-or-nothing condition but varies continuously over time \cite{Hiremath-Kumari-2014-SpringerPlus}.
In fact, the AMH offers several practical implications \cite{Lo-2004-JPM,Lo-2005-JIC}, including unstable relation of risk-premium, time-varying arbitrage opportunities, time-dependent and context-dependent performance of trading strategies, etc. Recent studies provide mounting empirical evidence supporting the AMH, most of which focus on testing time-varying return predictability as a measure of market efficiency
\cite{Kim-Shamsuddin-Lim-2011-JEF,Shi-Jiang-Zhou-2017-RAPS}.
Another strand of literature focuses on testing the time-varying performance of trading strategies implied in the AMH. The related method is to verify whether the risk-adjusted returns of trading strategies are time-varying and whether market conditions coincide with significantly good or poor performance. Empirical analysis includes the foreign exchange market \cite{Neely-Weller-Ulrich-2009-JFQA,Neely-Weller-2013-JBF}) and stock indexes including the US market \cite{Taylor-2014-JBF}, the Asian market \cite{Todea-Ulici-Silaghi-2009-RFB}, the Russian market \cite{Kinnunen-2013-EMR}, the Finnish market \cite{Patari-Vilska-2014-AE}, and the Romanian market
\cite{Anghel-2015-IJFR}. These results reveal the marked evidence in accordance with the AMH.

Our work contributes to the literatures on the AMH by testing several practical implications from the AMH within the framework of two renowned market anomalies, i.e., cross-sectional momentum and contrarian effects (hereafter, CSMOM and CSCON).
In fact, they are found to be ubiquitous in many asset classes as well as in most regions around the world. Although most earlier works about CSMOM and CSCON effects focus on the U.S. market \cite{DeBondt-Thaler-1985-JF,Jegadeesh-Titman-1993-JF,Jegadeesh-Titman-2001-JF,Gutierrez-Kelley-2008-JF}, a large body of studies find evidence of CSMOM or CSCON effects in other regions or markets, including the UK \cite{Hon-Tonks-2003-JMFM,Galariotis-Holmes-Ma-2007-JMFM}, Japan \cite{Chou-Wei-Chung-2007-JEF,Asness-2011-JPM}, Australia \cite{Demir-Muthuswamy-Walter-2004-PBFJ}, China \cite{Kang-Liu-Ni-2002-PBFJ,Wang-Chin-2004-PBFJ,Naughton-Truong-Veeraraghavan-2008-PBFJ,Pan-Tang-Xu-2013-PBFJ,Shi-Jiang-Zhou-2015-PLoS1}, etc.
Moreover, the CSMOM and CSCON effects are also found in many asset classes \cite{NovyMarx-2012-JFE,Asness-Moskowitz-Pedersen-2013-JF}, including stocks \cite{Shi-Jiang-Zhou-2015-PLoS1}, funds \cite{Grinblatt-Titman-Wermers-1995-AER,Carhart-1997-JF}, currency \cite{Kho-1996-JFE,Nitschka-2010-GER,Menkhoff-Sarno-Schmeling-Schrimpf-2012-JFE}, and commodity \cite{Erb-Harvey-2006-FAJ,Miffre-Rallis-2007-JBF}.
In addition, the frequently employed procedure to construct zero-cost arbitrage CSMOM or CSCON portfolios via buying (selling) winner and selling (buying) loser, enlightens us to test one of practical implications in the AMH that arbitrage opportunities exist from time to time \cite{Lo-2004-JPM}.

In this work, we focus on the Chinese stock market. Compared with mature financial markets, the Chinese stock market is relatively young. Some idiosyncratic phenomena characterize the Chinese market, including the less transparent information environment
at the market level and the firm level and a larger proportion of irrational individual investors, etc. We wonder whether the Chinese market is gradually improving to be efficient or evolving in the way similar to the descriptions in the AMH. In addition, few researches on the AMH have paid attention to the Chinese market. Our study aims to provide more empirical evidence of Chinese market to the literatures on the AMH. Specifically, we investigate if the relation of risk and reward in the Chinese stock market is time-varying within the framework of CSCON (CSMOM) effect and conduct a study about the development of arbitrage opportunities through examining the performance of CSCON portfolios. It also should be noted that moving window analysis is adopted to conduct the research in our work, which is argued to provide evidence with more detailed dynamic information than subperiod analysis does.
Finally, the study is also carried out to investigate the relation between the performance of CSCON portfolios and several market conditions to further test if the CSMOM and CSCON effects are context-dependent.

The rest of this paper is organized as follows. Section \ref{S1:Methods} and section \ref{S1:Data} describes the methodology and the data employed in the study, respectively. Section \ref{S1:Results} reports the empirical results. Section \ref{S1:Conclusion} concludes.

\section{Methodology}
\label{S1:Methods}


Following the common method to investigate the CSMOM and CSCON effects \cite{Jegadeesh-Titman-1993-JF,Jegadeesh-Titman-2001-JF,Kang-Liu-Ni-2002-PBFJ,Naughton-Truong-Veeraraghavan-2008-PBFJ,Pan-Tang-Xu-2013-PBFJ,Shi-Jiang-Zhou-2015-PLoS1}, we construct $J-K$ portfolios of losers and winners based on CSMOM (CSCON) strategies. Parameters $J$ and $K$ represent the lengths of estimation period and holding period, respectively.
Specifically, for the current month $t$, individual stocks are ranked according to their historical performance during the previous $J$ months. The winner refers to the decile group with the highest past average return and the loser is the decile group with the lowest past average return. By longing the loser (winner) and shorting the winner (loser), the CSCON (CSMOM) portfolio is constructed at the beginning of each month during the whole sample period. We evaluate the performance of loser, winner and CSCON (CSMOM) portfolios during the subsequent $K$ months. There is a CSCON (CSMOM) effect when the loser portfolios have better (worse) performance than the winner portfolios.
For the convenience, we use CON$(J,K)$ to denote the $J-K$ CSCON portfolios.

In addition, as argued in Ref.~ \cite{Lehmann-1990-QJE,Ball-Kothari-Wasley-1995-JF,Conrad-Gultekin-Kaul-1997-JBES}, the bid-ask spread, non-synchronous trading as well as the lack of liquidity would enlarge the CSMOM and CSCON effects. To avoid the biased results, we follow the common approach to skip one month between estimation period and the holding period \cite{Jegadeesh-Titman-1993-JF,Shi-Jiang-Zhou-2015-PLoS1}.

It should be emphasized that we follow the more practical method from Ref.~\cite{Galariotis-Holmes-Ma-2007-JMFM} to obtain the buy-and-hold return series rather than exactly follow the process from Ref.~\cite{Jegadeesh-Titman-1993-JF,Jegadeesh-Titman-2001-JF}, in which they consider the overlapping portfolios and rebalance the portfolio at each month. According to Ref.~\cite{Jegadeesh-Titman-1993-JF,Jegadeesh-Titman-2001-JF}, if the holding period is $K$ months, the portfolios at the time $t$ should contain the portfolio formed at time $t$ and the portfolios constructed in former $K-1$ months, and the average return at time $t$ is equally-weighted return of all portfolios at same time. Instead, we consider the buy-and-hold return series of portfolios constructed during the same period. To be specific, $K$-month buy-and-hold return is obtained by longing or shorting the portfolio formed at the month $t$ and then held for $K$ months. We test if the $K$-month return series significantly deviate from zero. Statistically significant positive (negative) return for CSCON portfolios indicates significant CSCON (CSMOM) effect.

\section{Data sets}
\label{S1:Data}

In the investigation of the risk-premium relation and arbitrage opportunities in the Chinese stock market through studying the performance of CSCON portfolios during moving time periods, we retrieve the monthly data from the RESSET database, including the dividend-adjusted and split-adjusted monthly returns for all A-share individual stocks and the monthly Fama-French three factors, which are constructed strictly following Ref.~\cite{Fama-French-1993-JFE}.

\begin{table}[!ht]
\centering
\caption{Basic information of stock exchanges investigated in this work.}
   \label{TB:DS:BBG}
   \begin{tabular}{cccccccccc}
   \hline
  {\textit{Stock exchange}} & {\textit{Country}} & {\textit{Start date}} & {\textit{End date}}  & {\textit{\# of stocks}} \\
   \hline
 {ASX} & {Australia} & {08/01/1968} & {03/11/2015} & {1889}  \\
 {HKEX} & {China} & {04/04/1986} & {03/11/2015} & {1770}  \\
 {KSE} & {Korea} & {01/07/1975} & {03/10/2015} & {754}  \\
 {LSE} & {UK} & {01/07/1986} & {03/11/2015} & {1703}  \\
 {NYSE} & {USA} & {01/04/1968} & {03/10/2015} & {1587}  \\
 {PSE} & {France} & {01/05/1977} & {03/10/2015} & {955}  \\
 {SHSE} & {China} & {12/31/1990} & {06/30/2015} & {1097}  \\
 {SZSE} & {China} & {12/10/1990} & {06/30/2015} & {1757}  \\
 {TSE} & {Tokyo} & {09/12/1974} & {03/11/2015} & {3467}  \\
 {TSX} & {Canada} & {01/06/1975} & {03/10/2015} & {911}  \\
 {TWSE} & {China} & {01/05/1991} & {03/10/2015} & {856}  \\
   \hline
   \end{tabular}
\begin{flushleft}
\end{flushleft}
\end{table}

Based on CSCON and CSMOM effects, we also study the arbitrage opportunities in other major stock exchanges. We utilize the monthly data of individual common stocks listed on some major stock exchanges in the world, which are downloaded from Bloomberg, covering the period from the time with earliest records in the database to March 2015. The data sets include dividend-adjusted and split-adjusted daily closing prices of individual stocks. We randomly select one representative for stock markets in different regions. The exchanges consist of Hong Kong Stock Exchange (HKEX), Taiwan Stock Exchange (TWSE), Tokyo Stock Exchange (TSE), Korea Stock Exchange (KSE), Australian Securities Exchange (ASX), London Stock Exchange (LSE), Paris Stock Exchange (PSE), New York Stock Exchange (NYSE), and Toronto Stock Exchange (TSX). Table \ref{TB:DS:BBG} reports the basic information about the stock exchanges in this study.

The study is also conducted to investigate the relation between the performance of CSCON portfolios and market condition factors, which are calculated based on some data, including the monthly closing price of Shanghai Component Index (SHCI), daily return and trading volume in RMB of all A-share individual stocks.

\section{Empirical results}
\label{S1:Results}

\subsection{Time-varying risk-premium relation}

As argued in Ref.~\cite{Lo-2004-JPM}, relative size and preferences of different traders and other factors, including regulatory environment and tax laws, lead to the unstable relation of risk-premium. In this section, using moving window approach with the window size being $5$ years, we intend to verify if there is an unstable relation between risk and reward within the framework of CSMOM or CSCON effect. Specifically, from the perspective of CSMOM or CSCON effect, the fluctuation of risk-adjusted returns for CSCON (CSMOM) portfolios and time-varying loadings on risk factors could provide evidence for the unstable relation between risk and reward. We employ two popular models, including the Capital Asset Pricing Model (CAPM) and the Fama-French three-factor model (FFTM), as described by Eq.~(\ref{Eq:CAPM}) and Eq.~(\ref{Eq:FFTM}), respectively. $ER_{t}$ denotes the return of CSCON portfolio at month $t$. $R_{\rm{MKT}}$, $R_{\rm{SMB}}$ and $R_{\rm{HML}}$ respectively represent the market factor, size factor and value factor from Ref.~\cite{Fama-French-1993-JFE}.
\begin{equation}
\centering
ER_{t}=\alpha+\beta_{\rm{MKT}}R_{{\rm{MKT}},t}+\varepsilon_{t}, ~~~\varepsilon_{t} \thicksim N(0,\sigma^{2}_{\varepsilon}),
\label{Eq:CAPM}
\end{equation}
\begin{equation}
\centering
ER_{t}=\alpha+\beta_{\rm{MKT}}R_{{\rm{MKT}},t}+\beta_{\rm{SMB}}R_{{\rm{SMB}},t}+\beta_{\rm{HML}}R_{{\rm{HML}},t}+\varepsilon_{t}, ~~~\varepsilon_{t} \thicksim N(0,\sigma^{2}_{\varepsilon}),
\label{Eq:FFTM}
\end{equation}

The empirical results of Ref.~\cite{Shi-Jiang-Zhou-2015-PLoS1} suggest a robust CSCON effect in Chinese stock market during the period 1997-2012. They also conclude that the performance of CSCON portfolios can be enhanced by prolonging the estimation period $J$, while it is insensitive to the holding horizons $K$. Hence, we consider six kinds of CSCON portfolios with different estimation periods $J$ belonging to $\{1,12,24,36,48,60\}$ month(s), while the holding period $K$ is fixed to $1$ month. The CSCON portfolios are constructed in terms of all A-share individual stocks in the Chinese stock market. Table \ref{TB:CON} reports the performance of CSCON portfolios for the period from 1990 to 2014, unveiling a long-term CSCON effect. We also provide the results for the period from 1997 to 2014, indicating the presence of both the short-term and long-term CSCON effects in China. These findings are consistent with Ref.~\cite{Shi-Jiang-Zhou-2015-PLoS1}. It is evident that the CSCON portfolios exhibit different performance during different time periods, and we will take a further investigation in the next section.

\setlength\tabcolsep{2pt}
\begin{table}[!ht]
  \centering
  \caption{{Raw returns of the CSCON portfolios constructed with different estimation periods $J$ and held in a fixed holding period $K=1$ month \cite{Shi-Jiang-Zhou-2015-PLoS1}. The \textit{t}-statistics presented in the parentheses are adjusted for heteroscedasticity and autocorrelation based on \cite{Newey-West-1987-Em}. Superscripts * and **, denote the significance at 5\% and 1\% levels, respectively.}}
  \label{TB:CON}
   \begin{tabular}{cccccccccccccccccccc}
   \hline
    && \multicolumn{2}{c}{\textit{ J=1}} && \multicolumn{2}{c}{\textit{12}} && \multicolumn{2}{c}{\textit{24}} && \multicolumn{2}{c}{\textit{36}} &&\multicolumn{2}{c}{\textit{48}} && \multicolumn{2}{c}{\textit{60}} \\
   \cline{3-4} \cline{6-7} \cline{9-10} \cline{12-13} \cline{15-16} \cline{18-19}
    period && \textit{R} & $t$-stat &&\textit{R} & $t$-stat && \textit{R} & $t$-stat && \textit{R} & $t$-stat && \textit{R} & $t$-stat && \textit{R} & $t$-stat \\
   \hline
 {1990-2014} &~~~& 0.008 & (1.91)$^{~~~}$ && 0.004 & (0.80)$~~~$ && 0.009 & (1.51)$~~~$ && 0.020 & (2.64)$^{**~}$ && 0.013 & (2.50)$^{*~~}$ && 0.015 & (2.46)$^{*~~}$\\
 {1997-2014} &~~~& 0.008 & (3.04)$^{**~}$ && 0.004 & (0.88)$~~~$ && 0.010 & (1.98)$^{*~~}$ && 0.014 & (2.72)$^{**~}$ && 0.016 & (2.79)$^{**~}$ && 0.017 & (3.04)$^{**~}$ \\
   \hline
   \end{tabular}
\end{table}

We conduct the study on the relation of risk and reward through examining the performance of aforementioned $6$ CSCON portfolios under the CAPM and the FFTM. The results are reported in Table \ref{TBS:Regression} for different estimation periods in the whole sample period 1990-2014. The regression slopes $\beta_{\rm{MKT}}$, $\beta_{\rm{SMB}}$ and $\beta_{\rm{HML}}$ are the corresponding loadings of the market factor $R_{\rm{MKT}}$, the size factor $R_{\rm{SMB}}$ and the value factor $R_{\rm{HML}}$.

\setlength\tabcolsep{0.5pt}
\begin{table}[!ht]
\centering
  \caption{{Adjusted returns of CSCON portfolios (the values of $\alpha$) under the CAPM and the Fama-French three-factor model as well as the loadings on different risk factors (the values of $\beta_{\rm{MKT}}$, $\beta_{\rm{SMB}}$, $\beta_{\rm{HML}}$). The superscripts * and ** denote the significance at 5\% and 1\% levels, respectively.}}
   \label{TBS:Regression}
   \begin{tabular}{ccccccccccccccccc}
   \hline
    && \multicolumn{4}{c}{\textit{CAPM}} && \multicolumn{8}{c}{\textit{Fama-French three-factor model}}\\
   \cline{3-6} \cline{8-15}
    && \multicolumn{2}{c}{$\alpha$} & \multicolumn{2}{c}{$\beta_{\rm{MKT}}$} &&  \multicolumn{2}{c}{$\alpha$} & \multicolumn{2}{c}{$\beta_{\rm{MKT}}$}  & \multicolumn{2}{c}{$\beta_{\rm{SMB}}$} & \multicolumn{2}{c}{$\beta_{\rm{HML}}$}\\
   \hline
 {$J=1$} &~~~~~&  0.008 & (1.48)$~~~$ & 0.123 & (3.03)$^{**~}$ &~~~~~& 0.009 & (1.66)$^{~~~}$ & 0.129 & (3.12)$^{**~}$ & -0.125 & (-1.53)$~~~$ & 0.030 & (0.35)$~~~$ \\
 {$J=12$} &~~~~~& 0.004 & (0.72)$~~~$ & 0.030 & (0.66)$~~~$ &~~~~~& 0.008 & (1.33)$~~~$ & 0.048 & (1.05)$~~~$ & -0.402 & (-4.49)$^{**~}$ & 0.083 & (0.89)$~~~$ \\
 {$J=24$} &~~~~~& 0.007 & (1.35)$~~~$ & 0.160 & (3.54)$^{**~}$ &~~~~~& 0.003 & (0.48)$~~~$ & 0.142 & (3.16)$^{**~}$ & 0.354 & (3.58)$^{**~}$ & 0.104 & (1.09)$~~~$ \\
 {$J=36$} &~~~~~& 0.017 & (2.56)$^{*~~}$ & 0.408 & (7.36)$^{**~}$ &~~~~~& 0.003 & (0.52)$~~~$ & 0.362 & (7.60)$^{**~}$ & 0.935 & (8.55)$^{**~}$ & 0.701 & (6.10)$^{**~}$ \\
 {$J=48$} &~~~~~& 0.013 & (2.48)$^{*~~}$ & 0.033 & (0.54)$~~~$ &~~~~~& 0.005 & (1.07)$~~~$ & -0.037 & (-0.67)$~~~$ & 0.683 & (6.92)$^{**~}$ & 0.422 & (3.77)$^{**~}$ \\
 {$J=60$} &~~~~~& 0.015 & (2.55)$^{*~~}$ & 0.003 & (0.05)$~~~$ &~~~~~& 0.008 & (1.47)$~~~$ & -0.057 & (-0.90)$~~~$ & 0.620 & (5.72)$^{**~}$ & 0.269 & (2.07)$^{*~~}$ \\ \hline
   \end{tabular}
\end{table}

We first check the results for the CAPM.
In general, compared with the results for raw returns of CSCON portfolios during full sample period in Table \ref{TB:CON}, the risk-adjusted returns (i.e., the values of $\alpha$) of CSCON portfolios in Table~\ref{TBS:Regression} are only slightly reduced or remain unchanged. The introduction of the market factor $R_{\rm{MKT}}$ does not cause substantial reduction in the returns. The risk-adjusted returns are positive and statistical significant for the CSCON portfolios with longer estimation periods $J\geq 36$. However, the loadings on the market factor are statistically significant for $J=1$, $24$, and $36$. As a whole, the CAPM does not have strong power to explain the returns of CSCON strategies.
The risk-adjusted returns in moving windows vary over time and are sensitive to the estimation period $J$. When $J=1$, the risk-adjusted returns are significantly positive in four windows (1996-2000, 1997-2001, 1998-2002, and 2001-2005), which are greater than the raw (and the risk-adjusted as well) return in the whole sample. When $J=12$ and 24, there are two windows (2007-2011 and 2008-2012) with significantly positive risk-adjusted returns. With the increase of $J$, there are more moving windows around 2007-2012 with significantly positive risk-adjusted returns. The loadings $\beta_{\rm{MKT}}$ on the market factor are significantly positive in some moving windows for certain $J$ values, significantly negative in some other windows, or insignificant for the rest cases.

We next investigate the results for the FFTM.
For short estimation periods ($J\leq12$), the risk-adjusted returns of the whole sample in Table~\ref{TBS:Regression} are higher than the raw returns of the CSCON portfolios reported in Table \ref{TB:CON}. It means that the FFTM does not exhibit stronger explanative power for short estimation periods. For larger estimation periods ($J\geq24$), the risk-adjusted returns are substantially smaller and become statistically insignificant, indicating that the FFTM is more favorable than the CAPM in explaining the CSCON effect. Overall, the loadings $\beta_{\rm{MKT}}$ on the market factor are significantly positive for short estimation periods, the loadings $\beta_{\rm{SMB}}$ on the size factor are significantly positive for long estimation periods ($J\geq24$), and the loadings $\beta_{\rm{HML}}$ on the value factor are significantly positive for longer estimation periods ($J\geq36$). For $J=12$, $\beta_{\rm{SMB}}$ is significantly negative. The results in moving windows are again mixed. Significant and insignificant risk-adjusted returns are observed in different windows. This implies that there might be a time-varying CSMOM or CSCON effect in the Chinese stock market.
In the first a few windows, the loadings $\beta_{\rm{MKT}}$ on the market factor are statistically positive for some estimation periods, while in the sequel windows, the loadings $\beta_{\rm{MKT}}$ become statistically negative for some estimation periods. The loadings $\beta_{\rm{SMB}}$ on the size factor are significantly negative in the early years for small $J$ values and are significantly positive in later years for most $J$ values. In contrast, the loadings $\beta_{\rm{HML}}$ are insignificant in all moving windows for $J=1$ and significantly positive in most moving windows for $J\geq12$.

In summary, both the risk-adjusted returns for CSCON portfolios and loadings on different risk factors are time-varying. According to Ref.~\cite{Fama-1970-JF,Fama-1991-JF}, the EMH suggests that the market would keep being efficient or at least is becoming more efficient. However, when $J>1$, the risk-adjusted returns during earlier time windows are not statistically significant for both models, while those for later time windows (mainly from 2005 to 2014) the returns turn out to be statistically significant. This indicates that the market does not continue to be more effective. Rather, the efficiency of the Chinese stock market changes over time. It is evident that these findings are in accordance with the description about the market efficiency under the AMH framework \cite{Lo-2004-JPM}.

Our findings also cast doubts on the usual process adopted in most of the previous literature, in which the constant loadings on risk factors are implied by risk adjustment through performing time series regression upon full sample period. It is flawed that the dynamics of factor loadings is not included in conventional process to perform the risk adjustment. Theoretically, the loadings on risk factors are indeed time-varying and dependent on the frequency of CSMOM (CSCON) portfolios construction, as pointed out by Ref.~\cite{Wang-Wu-2011-JBF}. In the basis of the assumption that the returns for individual stocks are generated by the FFTM with constant loadings on risk factors, the monthly re-construction of CSMOM (CSCON) portfolios that is widely used in the literature requires different stocks selected for winner and loser month by month. And the loadings on risk factors for CSMOM (CSCON) portfolios are the average loadings for individual stocks, thus leading to time-varying loadings on risk factors.
On the other hand, it is intuitive that the winner (loser) portfolio is constructed by the stocks with generally higher (lower) loadings if the risk factor premium has positive returns, while the stocks are selected for winner (loser) mostly with the lower (higher) loadings of the risk factor if the risk factor premium has positive returns. Therefore, they conclude that the risk factors loadings co-vary with respective risk factor premium.
%

\subsection{Evolving arbitrage opportunities based on CSCON effect}

We now turn to test the hypothesis derived from the AMH that arbitrage opportunities appear and disappear over time by evaluating the performance of CSCON zero-cost arbitrage portfolios.

Specifically, we conduct the study on moving time windows. For the return time series $\{R_{J,K}^{i}\}_{1}^{T}$ obtained from CSCON portfolios, an initial sub-sample is created consisting of observations $\{R_{J,K}^{i}\}_{1}^{M}$ in the first moving window of size $M$ months. We then test whether the subsequence $\{R_{J,K}^{i}\}_{1}^{M}$ is significantly different from zero. Following Ref.~\cite{Newey-West-1987-Em}, we obtain the $t$-statistic adjusted for heteroscedasticity and autocorrelation. The window rolls forward by $a$ months to cover $\{R_{J,K}^{i}\}_{1+a}^{M+a}$ for re-checking the performance of the corresponding subsequence, and so forth until the end of the sample $\{R_{J,K}^{i}\}_{T-M+1}^{T}$. Following the majority of existing literatures, we set the value of $a$ as 12 months and $M$ as 60 months. We consider $11\times11=121$ CSCON portfolios with the estimation period $J$ and the holding period $K$ belonging to the set $\{1,6,12,18,24,30,36,42,48,54,60\}$.
For each time window, the CSCON portfolios can be divided into four groups according to their performance, namely, the portfolios having significantly positive returns (abbreviated as SP), the portfolios having significantly negative returns (abbreviated as SN), the portfolios having insignificantly positive returns (abbreviated as NSP), and the portfolios having insignificant negative returns (abbreviated as NSN). The significance level is 5\%.

\begin{figure}[!ht]
  \centering
\includegraphics[width=0.48\linewidth]{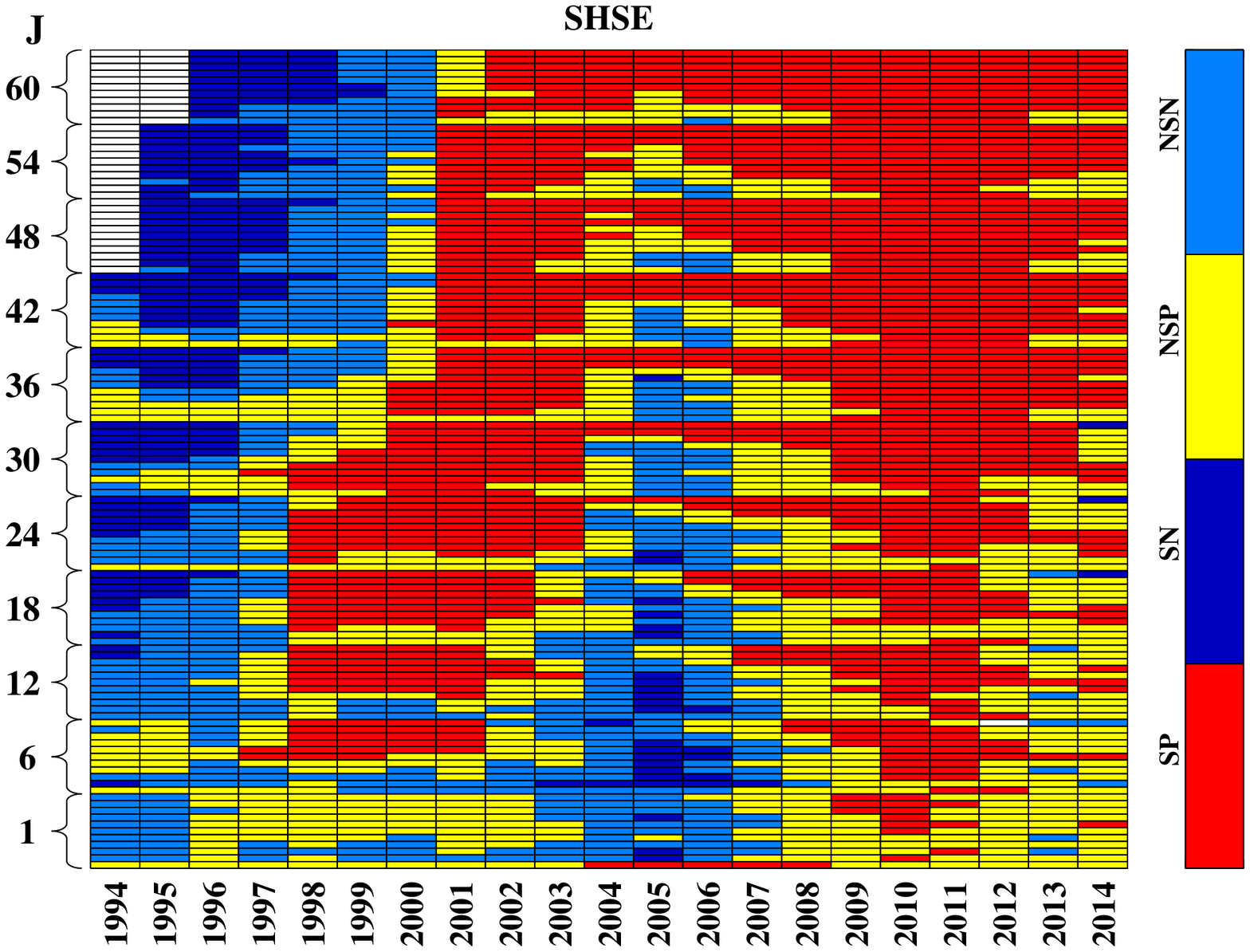}
\includegraphics[width=0.48\linewidth]{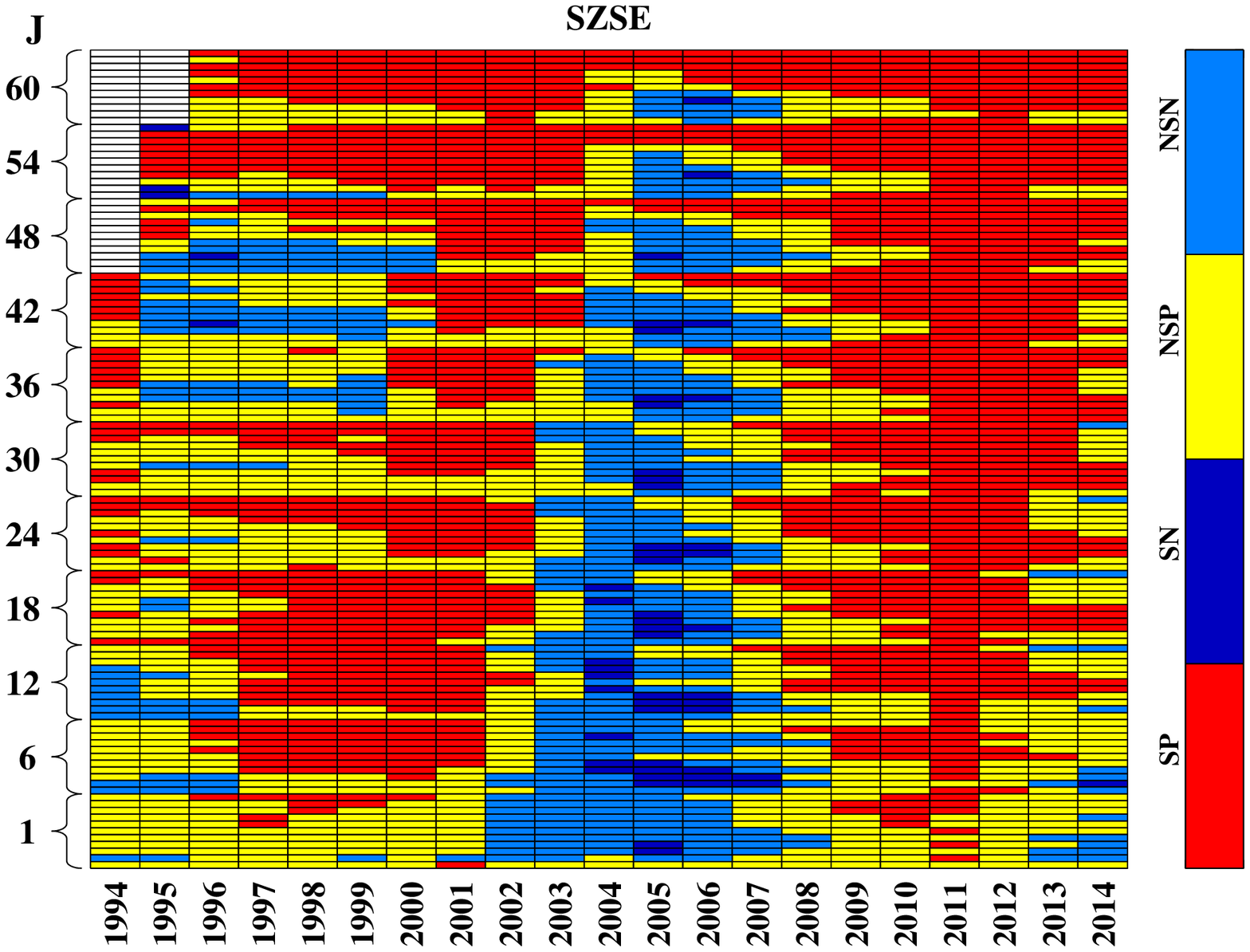}
  \caption{\label{Fig:RollingCON} (color online) Time-varying significance of the returns obtained from the CSCON portfolios in moving windows for the SHSE and SZSE. The vertical axis corresponds to $121$ CSCON portfolios with different estimations ($J$) and holding ($K$) horizons. The CSCON portfolios are constructed during the moving time intervals, which are presented on the horizontal axis. The years in the $x$-axis are the last years in the moving windows of five years.}
\end{figure}

The results are illustrated by Fig.~\ref{Fig:RollingCON}. The four distinct colors denote four groups formed according to the performance for $J-K$ CSCON portfolios. The changing of colors over time indicates that the arbitrage opportunities are indeed time-varying, which is ubiquitous for all stock exchanges. The grids without any color stand for no data available. Although the start dates for stocks are reported for some exchanges in Table \ref{TB:DS:BBG}, no sufficient data are available in the early moving windows due to long estimation periods or no sufficient stocks to construct portfolios. In each plot, we observe vertical strips, implying that the returns are less dependent on $J$ and $K$ than on time $t$. A closer scrutiny unveils a periodic pattern vertically, which reflects the fact that the returns are actually dependent of the holding horizons $J$ \cite{Shi-Jiang-Zhou-2015-PLoS1}.

We find that these two plots share striking similarity, except for the left-top corner (early years, large estimation periods). In the early years from 1994 to 1997, most SHSE CSCON portfolios have negative returns and some of them are significant, while most SZSE CSCON portfolios have positive returns and quite a few are significant. In other words, from 1994-1997, the SHSE stocks exhibit a momentum effect, while the SZSE stocks exhibit a CSCON effect. From 1998 to 2000, the SHSE CSCON portfolios are insignificantly negative for large $J$'s and significantly positive for small $J$'s, while the SZSE CSCON portfolios are basically positive. Since 2001, both exchanges show a CSCON effect with a weakening around 2005.

Our findings provide convincing evidence that arbitrage opportunities based on CSCON portfolios would evolve over time and might be dependent on investment environments. In Ref.~\cite{Shi-Jiang-Zhou-2015-PLoS1}, the investigations of the existence of a CSMOM or CSCON effect are conducted in different stock markets and during different sample periods. According to Fig.~\ref{Fig:RollingCON}, it is not surprising that previous studies report somewhat contradictory results. In other words, the market anomalies  evolve over time rather than remain a static state. Our analysis suggests that the AMH should be considered when we study the behaviors of market anomalies.

\begin{figure}[!ht]
  \centering
  \includegraphics[width=0.33\linewidth]{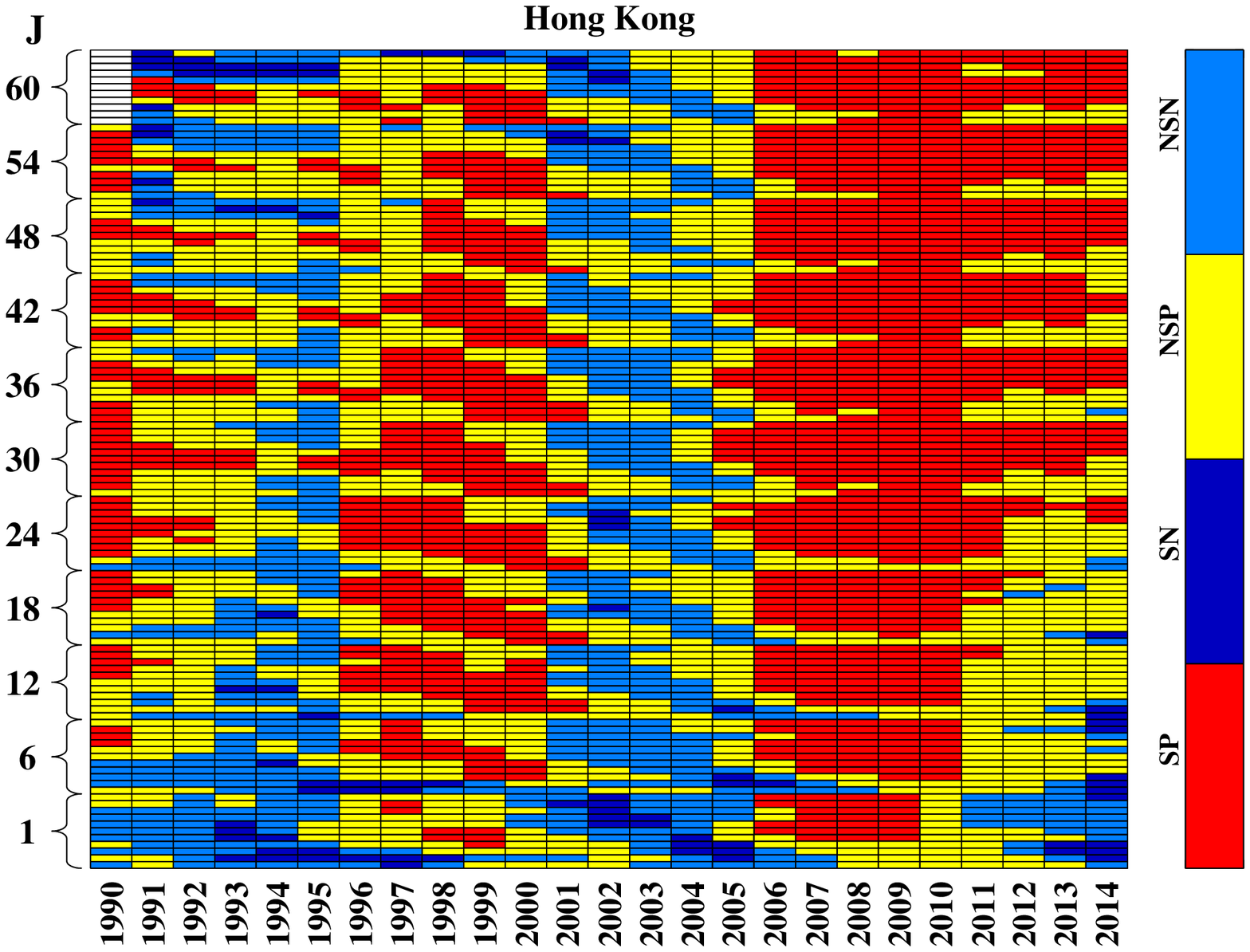}
  \includegraphics[width=0.33\linewidth]{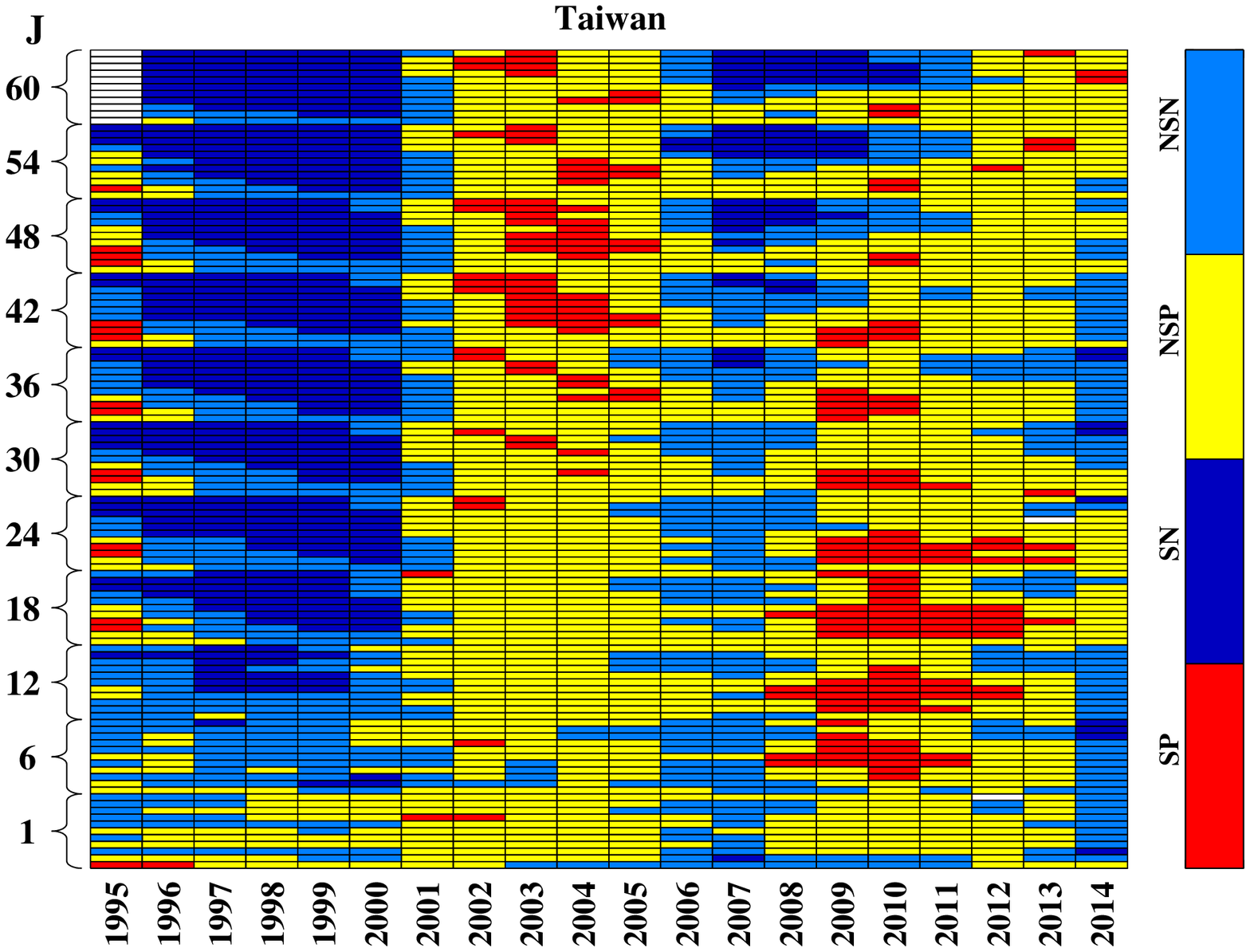}
  \includegraphics[width=0.33\linewidth]{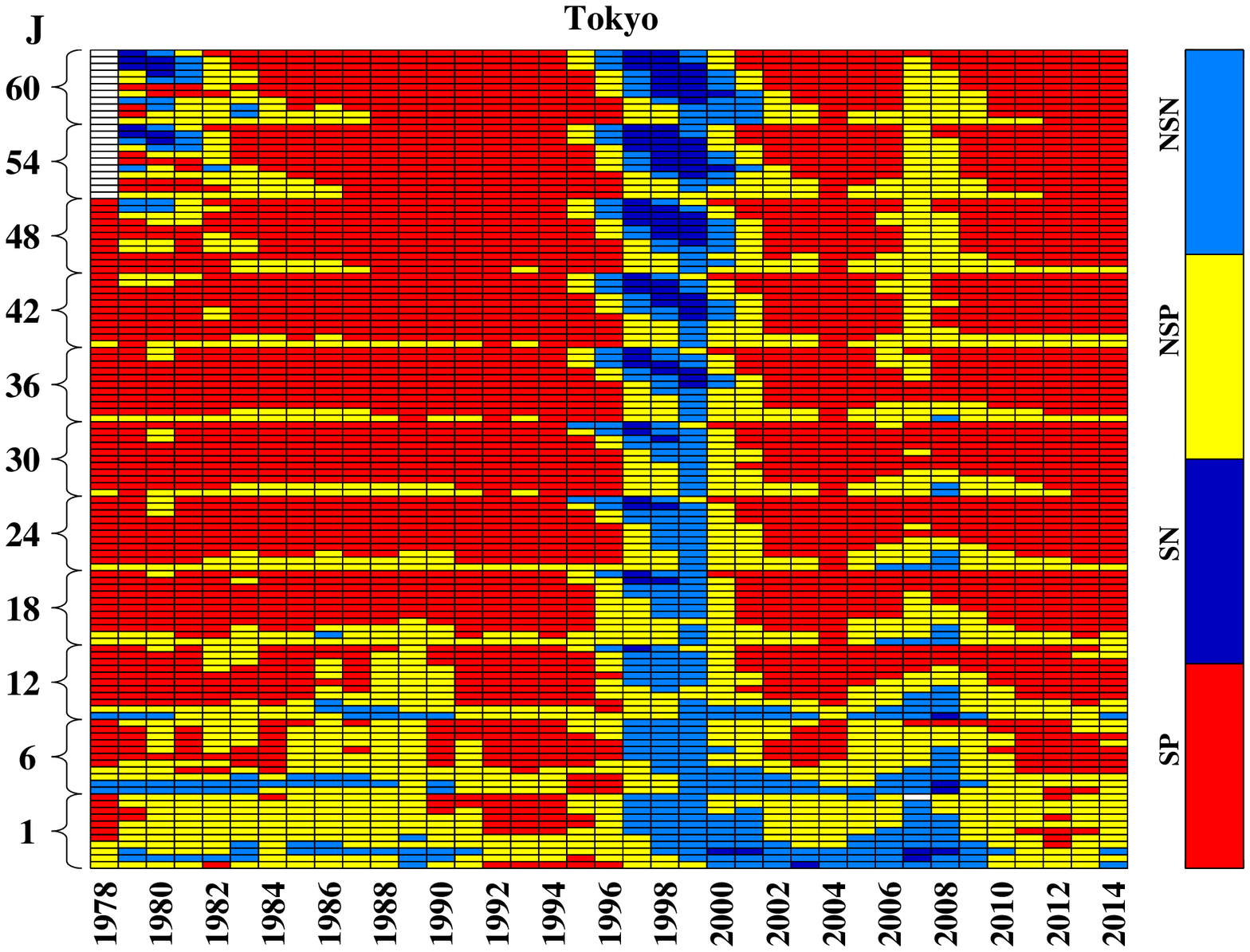}
  \includegraphics[width=0.33\linewidth]{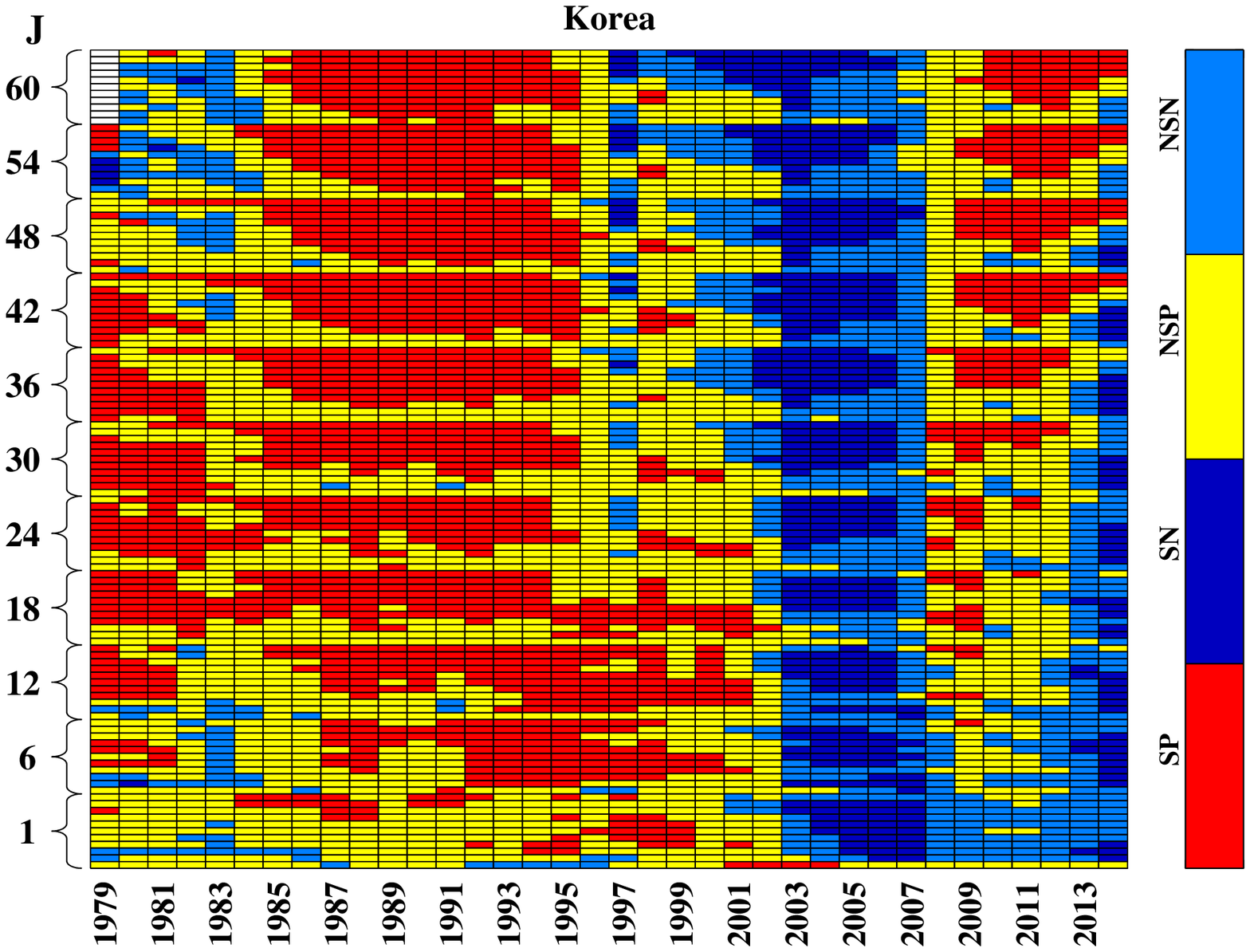}
  \includegraphics[width=0.33\linewidth]{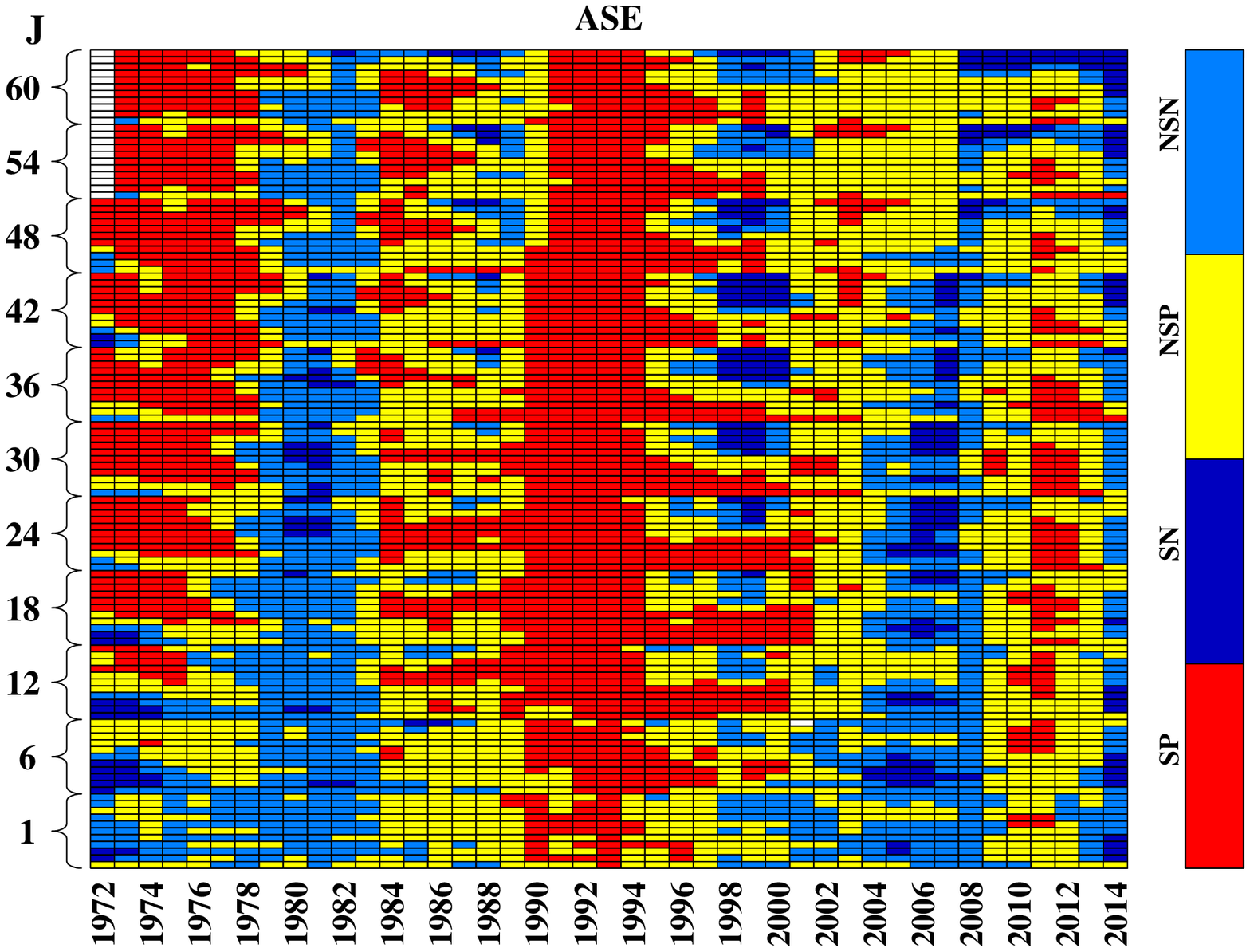}
  \includegraphics[width=0.33\linewidth]{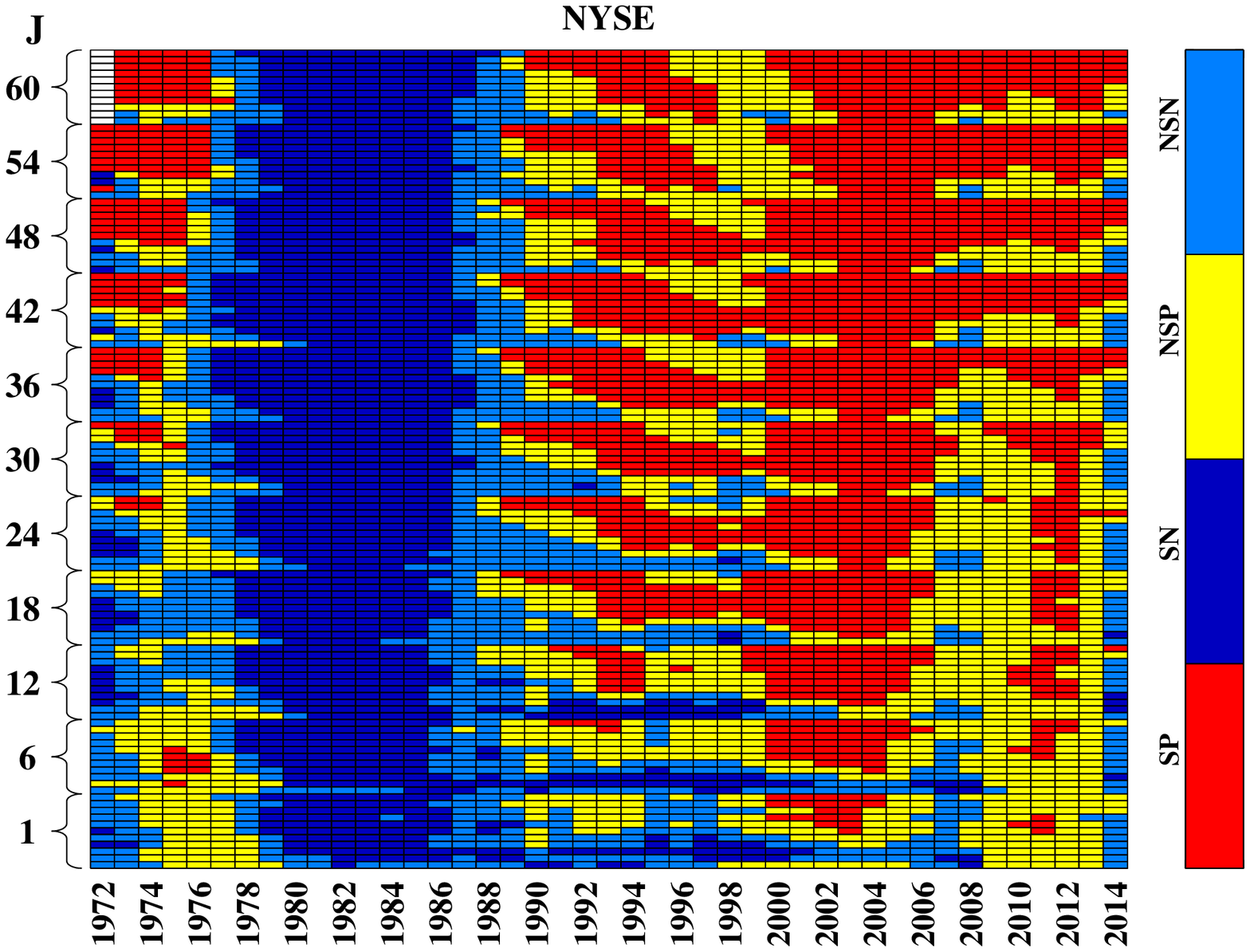}
  \includegraphics[width=0.33\linewidth]{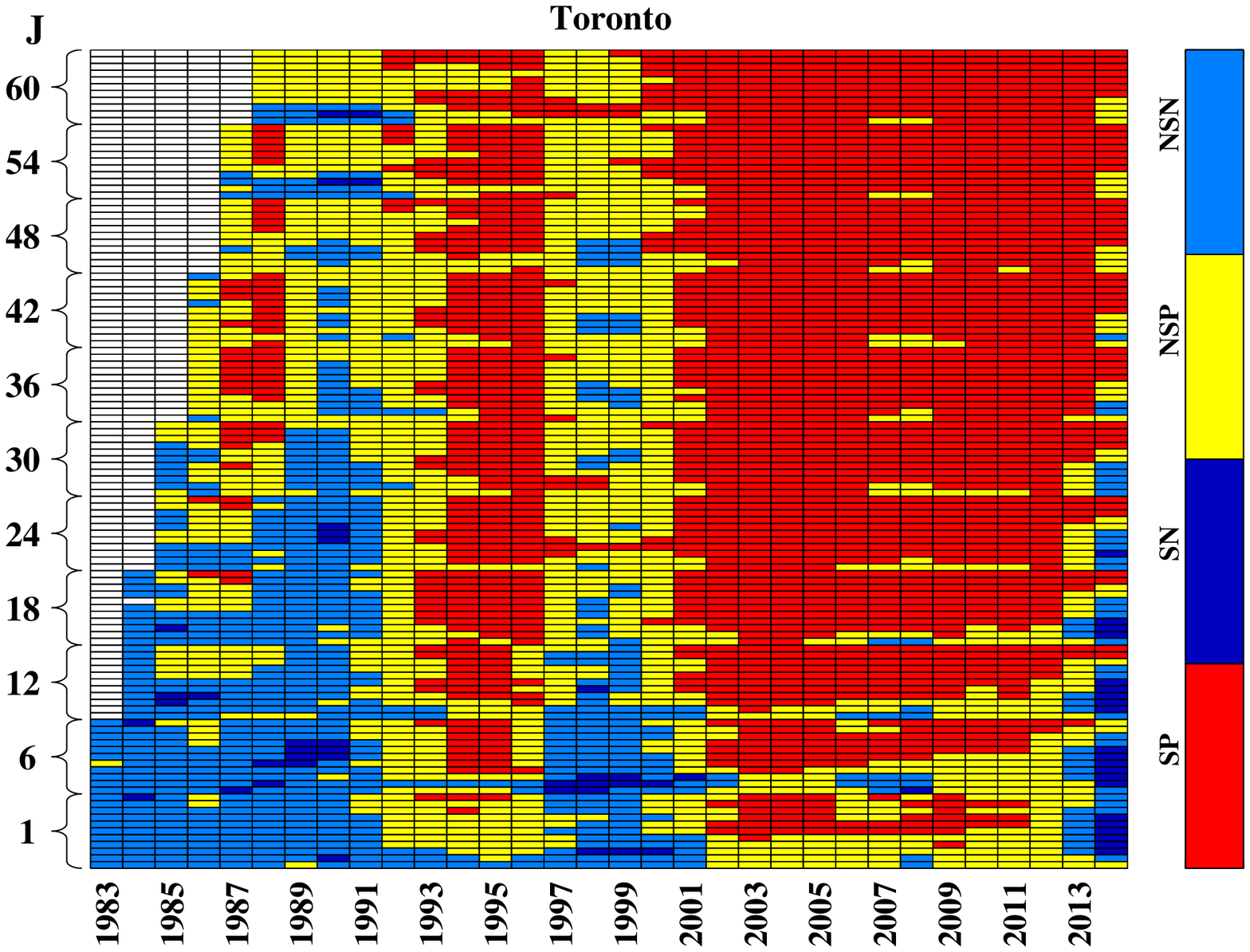}
  \includegraphics[width=0.33\linewidth]{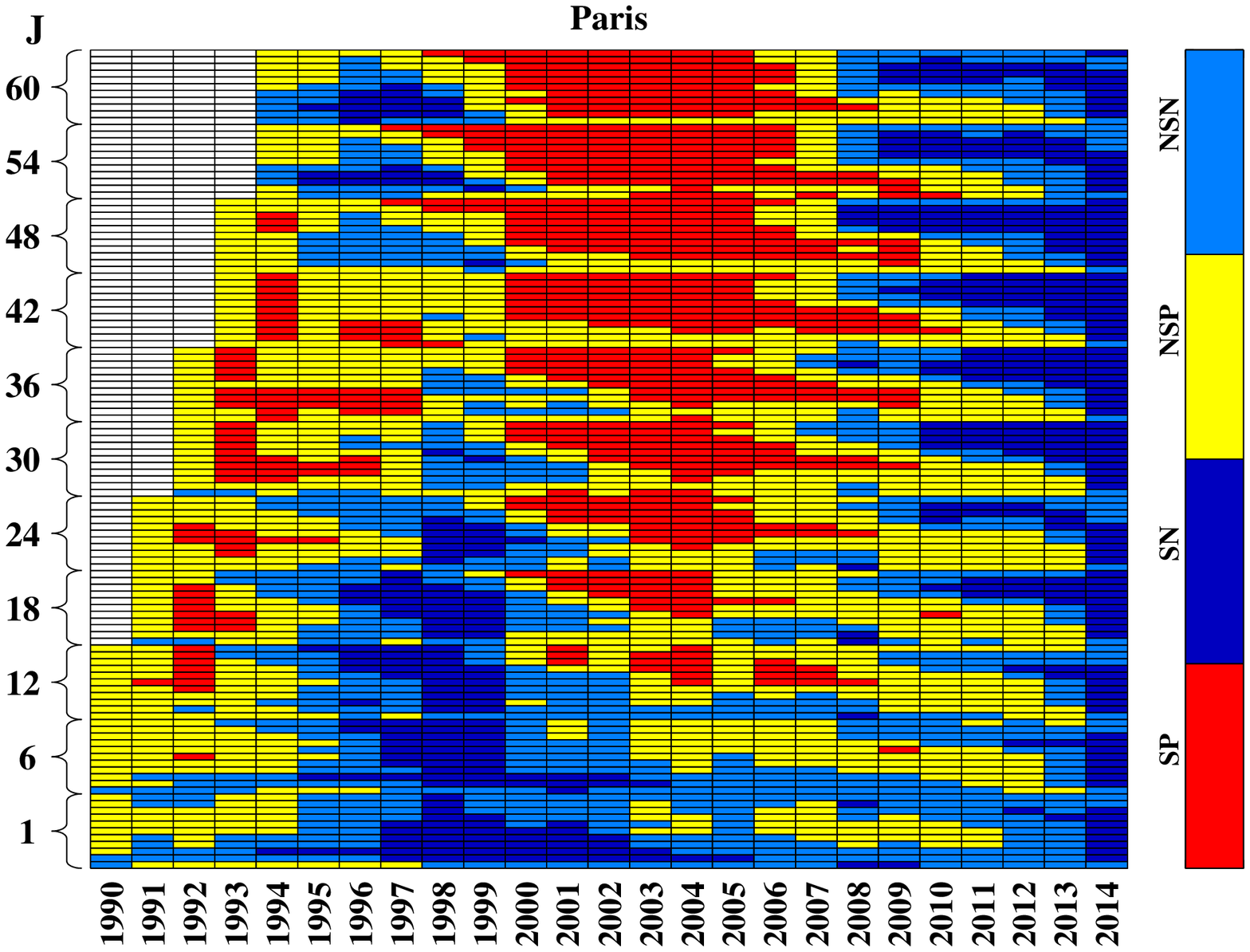}
  \includegraphics[width=0.33\linewidth]{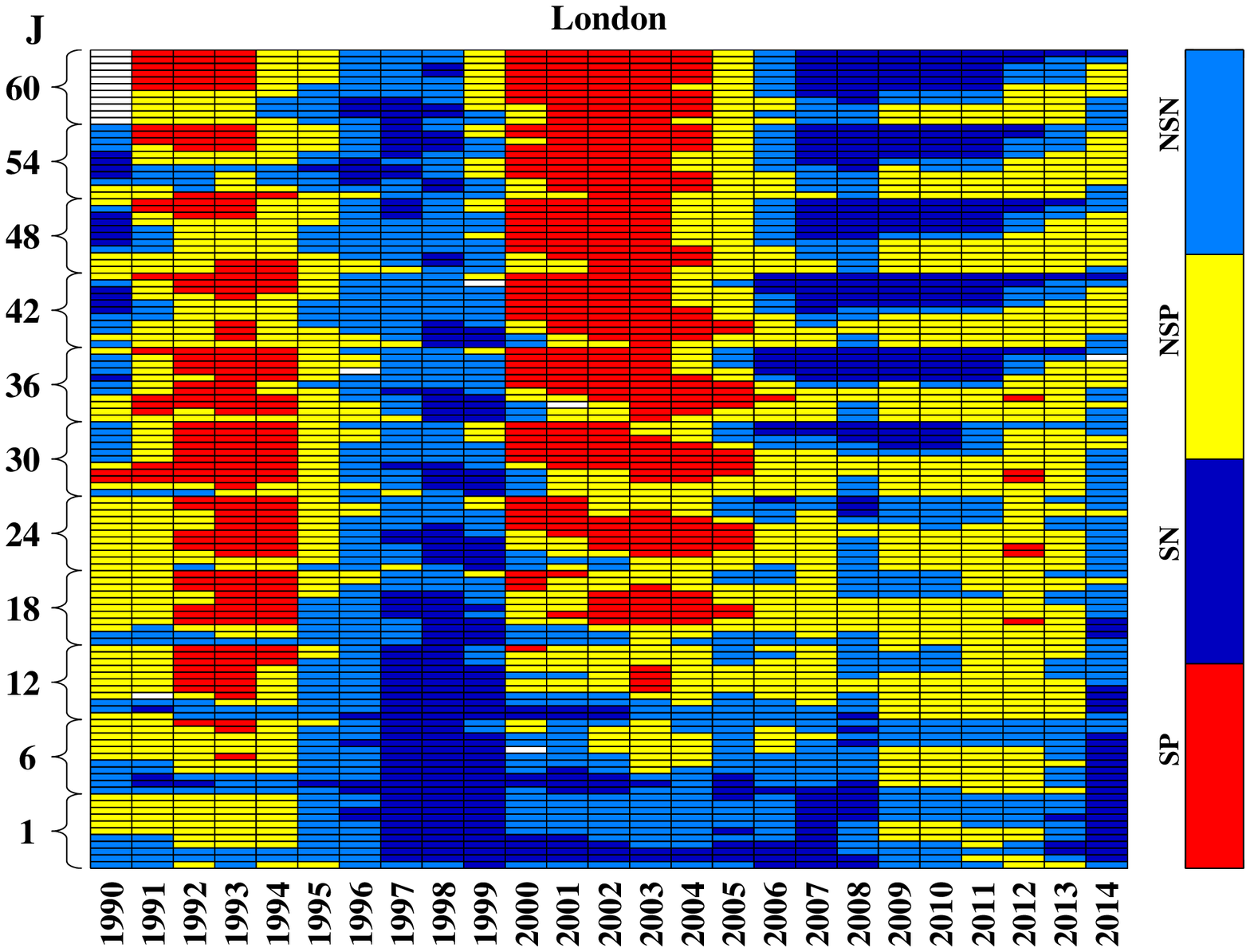}
  \caption{\label{Fig:RollingCON:Others} (color online) Time-varying significance of the returns obtained from the CSCON portfolios in moving windows for 9 major stock exchanges outside Mainland China. The vertical axis corresponds to $121$ CSCON portfolios with different estimations ($J$) and holding ($K$) horizons. The CSCON portfolios are constructed during the moving time intervals, which are presented on the horizontal axis. The years in the $x$-axis are the last years in the moving windows of five years.}
\end{figure}

We also perform the same analysis for 9 main stock exchanges outside Mainland China for comparison. The results are presented in Fig.~\ref{Fig:RollingCON:Others}. For the Hong Kong Stock Exchange, we observe an overall CSCON effect, which gradually weakens from 1990 to 1995, becomes strong from 1996 to 2000, disappears from 2001 to 2005, and appears again from 2006 to 2014. For the Taiwan Stock Exchange, we observe a mixture of CSMOM and CSCON effects in 1995, a momentum effect from 1996 to 2000, a weak CSCON effect from 2002 to 2005 for some $J$'s, a CSMOM effect in 2007 and 2008, and a CSCON effect afterwards. For the Tokyo Stock Exchange, a CSCON effect dominates except for large $J$'s from 1978 to 1982 and for all all $J$'s from 1997 to 2000. For other exchanges, we also observe the wax and wane of the CSCON and CSMOM effects.

\subsection{Context-dependent CSMOM and CSCON effects}

\begin{table}[!ht]
  \caption{This table reports the performances of the cross-sectional $J-K$ contrarian (CSCON) portfolios during the period with high and low level of different market conditions, including market state (State), market volatility (Volatility), market illiquidity (Illiquidity), and macroeconomic uncertainty (Uncertainty).
  Estimation period $J \in \{1,6,12,18,24,30,36,42,48,54,60\}$ month(s) and holding period $K=1$ month, which are presented respectively in first two columns. Monthly average returns of the CSCON portfolios during the whole sample period from 1990 to 2014 are also reported in third column, Raw~Return, against which the CSCON profitability for periods with high and low level of each market condition can compare.
  The sample period is December 1990 to  December 2014. Following Ref.~\cite{Newey-West-1987-Em}, $t$-statistics are adjusted for heteroscedasticity and autocorrelation. The superscripts * and ** denote the significance at 5\% and 1\% levels, respectively.}
   \label{TB:MarketCondition:CON}
   \centering
   \begin{tabular}{ccccccccccccccccccccc}
   \hline
   &&&&&& \multicolumn{3}{c}{State} && \multicolumn{3}{c}{Volatility} && \multicolumn{3}{c}{Illiquidity} && \multicolumn{3}{c}{Uncertainty}  \\
   \cline{7-9} \cline{11-13} \cline{15-17} \cline{19-21}
     $J$ && $K$ && Raw~Return && Up && Down && High && Low && High && Low && High && Low \\
   \hline
   \vspace{-3mm}\\
   1 && 1&& {0.0084$^{~~}$}  && {0.0097$^{~~}$} && {0.0057$^{~~}$} && {0.0115$^{~~}$}  && {\bf 0.0059$^{*~}$} && {0.0074$^{~~}$} && {0.0093$^{~~}$} && {0.0063$^{~~}$}&& {\bf 0.0105$^{*~}$}  \\
   6 && 1&& {0.0009$^{~~}$}  && {\bf 0.0104$^{*~}$} && {\bf -0.0113$^{*~}$} && {0.0069$^{~~}$}  && {-0.0041$^{~~}$} && {-0.0015$^{~~}$} && {0.0029$^{~~}$} && {-0.0021$^{~~}$}&& {0.0037$^{~~}$}  \\
   12 && 1&& {0.0044$^{~~}$}  && {0.0097$^{~~}$} && {-0.0077$^{~~}$} && {0.0139$^{~~}$}  && {-0.0035$^{~~}$} && {-0.0005$^{~~}$} && {0.0087$^{~~}$} && {0.0041$^{~~}$}&& {0.0047$^{~~}$}  \\
   18 && 1&& {0.0057$^{~~}$}  && {\bf 0.0141$^{*~}$} && {-0.0033$^{~~}$} && {0.0156$^{~~}$}  && {-0.0024$^{~~}$} && {-0.0003$^{~~}$} && {0.0110$^{~~}$} && {0.0002$^{~~}$}&& {0.0111$^{~~}$}  \\
   24 && 1&& {0.0089$^{~~}$}  && {\bf 0.0186$^{*~}$} && {-0.0001$^{~~}$} && {0.0191$^{~~}$}  && {0.0010$^{~~}$} && {-0.0021$^{~~}$} && {\bf 0.0183$^{*~}$} && {0.0021$^{~~}$}&& {\bf 0.0153$^{*~}$}  \\
   30 && 1&& {0.0150$^{~~}$}  && {\bf 0.0255$^{*~}$} && {0.0086$^{~~}$} && {0.0288$^{~~}$}  && {0.0047$^{~~}$} && {0.0042$^{~~}$} && {\bf 0.0241$^{**}$} && {0.0090$^{~~}$}&& {\bf 0.0203$^{**}$}  \\
   36 && 1&& {\bf 0.0204$^{**}$}  && {\bf 0.0263$^{*~}$} && {0.0121$^{~~}$} && {\bf 0.0409$^{*~}$}  && {0.0060$^{~~}$} && {0.0112$^{~~}$} && {\bf 0.0279$^{**}$} && {0.0205$^{~~}$}&& {\bf 0.0204$^{**}$}  \\
   42 && 1&& {\bf 0.0228$^{**}$}  && {\bf 0.0320$^{**}$} && {0.0103$^{~~}$} && {\bf 0.0484$^{**}$}  && {0.0058$^{~~}$} && {0.0177$^{~~}$} && {\bf 0.0267$^{**}$} && {\bf 0.0271$^{*~}$}&& {\bf 0.0194$^{*~}$}  \\
   48 && 1&& {\bf 0.0134$^{*~}$}  && {\bf 0.0261$^{**}$} && {-0.0031$^{~~}$} && {\bf 0.0292$^{**}$}  && {0.0036$^{~~}$} && {0.0050$^{~~}$} && {\bf 0.0196$^{*~}$} && {0.0094$^{~~}$}&& {\bf 0.0165$^{*~}$}  \\
   54 && 1&& {\bf 0.0151$^{**}$}  && {\bf 0.0301$^{**}$} && {-0.0041$^{~~}$} && {\bf 0.0320$^{**}$}  && {0.0052$^{~~}$} && {0.0103$^{~~}$} && {\bf 0.0183$^{*~}$} && {0.0055$^{~~}$}&& {\bf 0.0219$^{**}$}  \\
   60 && 1&& {\bf 0.0146$^{*~}$}  && {\bf 0.0290$^{**}$} && {-0.0042$^{~~}$} && {\bf 0.0284$^{*~}$}  && {0.0070$^{~~}$} && {0.0080$^{~~}$} && {\bf 0.0188$^{*~}$} && {0.0033$^{~~}$}&& {\bf 0.0221$^{**}$}  \\
   \hline
   \end{tabular}
\end{table}

The results above suggest the time-varying arbitrage opportunities or performance of CSCON strategies in the Chinese stock markets.
We thereby intend to examine the relation between CSCON (CSMOM) effects and some market condition factors that could characterize the investment context.

We carry out the studies from the following different perspectives. Firstly, we evaluate the performance of the $J-K$ CSCON portfolios formed by the Chinese A-share individual stocks during the period with both high and low level of specific market condition.
As with the samples in studying time-varying risk-premium relation, we take the CSCON portfolios with varying estimation period $J$ and fixed holding period $K$ as samples. $J \in \{1,6,12,18,24,30,36,42,48,54,60\}$ month(s) and $K=1$ month.
We employ several market condition factors that are frequently used in the previous literature
\cite{Taylor-2014-JBF}, including market state, market volatility, market illiquidity, and macroeconomic uncertainty.

On the other hand, in order to avoid the influence of common risk factors, we also turn to the pricing model with inclusion of dummy variables for market conditions to capture the more precise relation between the CSCON profitability and market conditions
\cite{Kim-Shamsuddin-Lim-2011-JEF,Taylor-2014-JBF}.
Specifically, the augmented FFTM is employed, as depicted by Eq.~(\ref{Eq:ThreeD}). $ER_t$ denotes the return of CSCON portfolio at month $t$. $R_{\rm{MKT}}$, $R_{\rm{SMB}}$ and $R_{\rm{HML}}$ respectively represent the market factor, size factor and value factor from Ref.~\cite{Fama-French-1993-JFE}. $D_{t}$ denotes the dummy variable of market condition, taking the value of unity (zero) if it is the high (low) level of market condition at month $t$.

\begin{equation}
\centering
ER_{t}=\alpha+\beta_{\rm{MKT}}R_{{\rm{MKT}},t}+\beta_{\rm{SMB}}R_{{\rm{SMB}},t}+\beta_{\rm{HML}}R_{{\rm{HML}},t}+\beta_{D}D_{t}+\varepsilon_{t}, ~~~\varepsilon_{t} \thicksim N(0,\sigma^{2}_{\varepsilon}),
\label{Eq:ThreeD}
\end{equation}

Market states consist of two trends, i.e., upward and downward. In fact, the state (or direction) of market is critically important to the profitability of CSMOM (CSCON) strategies \cite{Cooper-Gutierrez-Hameed-2004-JF}. Following Ref.~\cite{Cooper-Gutierrez-Hameed-2004-JF}, we define the upward (downward) trend of market state when the market performance over the previous 36-month is non-negative (negative). The monthly log return of the SHCI is employed as the proxy for market performance in our work. And its dummy variable at month $t$ equals to unity (zero) if it is the non-negative (negative) return of the SHCI over the previous 36 months.


Market volatility is defined as the conditional variance of $AR(1)$-$GJR$-$GARCH(1,1)$
\cite{Glosten-Jagannathan-Runkle-1993-JF}, 
as depicted by Eq.~(\ref{Eq:GJR}), where $R_{t}$ denotes the monthly log return of the SHCI. In Eq.~(\ref{Eq:GJR}),
the conditional mean and variance equations are described by the first and third equations, respectively.
$c$ and $\varepsilon_{t}$ are the constant and innovation terms, respectively. $\sigma_{t}$ represents the conditional standard deviation of $\varepsilon_{t}$. In variance equation, $\rm{I}[\cdot]$ is indicator function, taking the value of unity when $\varepsilon_{t-1}<0$ and zero otherwise. In other words, $\xi$ quantifies the difference between the effects of previous positive shocks ($\varepsilon_{t-1}>0$) and negative shocks ($\varepsilon_{t-1}<0$) on the volatility
\cite{Wang-Xie-Jiang-Stanley-2016-IREF}.
We adopt $AR(1)$-$GJR$-$GARCH(1,1)$ instead of $AR(1)$-$GARCH(1,1)$ from Ref.~\cite{Taylor-2014-JBF} on the ground that a surge of literatures find that $AR(1)$-$GJR$-$GARCH(1,1)$ could be more eligible to describe the dynamic of asymmetrical market volatility in the Chinese stock market.
The dummy variable of market volatility equals to unity (zero) in the high (low) volatility regime defined as periods in which market volatility is above (below) its median value.
\begin{equation}
\centering
\begin{aligned}
& R_{t}=c+R_{t-1}+\varepsilon_{t},\\
& \varepsilon_{t}=\sigma_{t}z_{t},\\
& \sigma^{2}_{t}=k+\gamma\sigma^{2}_{t-1}+\alpha\varepsilon^{2}_{t-1}+\xi
\rm{I}[\varepsilon_{t-1}<0]\varepsilon^{2}_{t-1},\\
\end{aligned}
\label{Eq:GJR}
\end{equation}

We follow Ref.~\cite{Amihud-2002-JFinM} to measure the market illiquidity, which is applied to all Chinese A-share stocks over a daily frequency. The results are averaged over all stocks and all days in each month to give a market-wide monthly frequency illiquidity estimate, as depicted by Eq. (\ref{Eq:Illiquidity}), where $R_{i,m,d}$ is the return of the stock $i$ on day $d$ of month $m$, $VOLD_{i,m,d}$ is the respective daily trading volume in RMB, $D_{i,m}$ is the number of days with available trading data and $N_{m}$ is the number of available stocks of month $m$. The high (low) market illiquidity regime is defined as periods in which the monthly market illiquidity is above (below) its median value. Correspondingly, the dummy variable equals to unity (zero) in the high (low) market illiquidity regime.
\begin{equation}
\centering
\begin{aligned}
& ILLIQ_{i,m}=\frac{1}{D_{i,m}}\sum_{t=1}^{D_{i,m}}\frac{|R_{i,m,d}|}{VOLD_{i,m,d}},\\
&
AILLIQ_{m}=\frac{1}{N_{m}}\sum_{t=1}^{N_{m}}ILLIQ_{i,m},\\
\end{aligned}
\label{Eq:Illiquidity}
\end{equation}

According to Ref.~\cite{Taylor-2014-JBF}, macroeconomic uncertainty for the US market is defined as the conditional variance associated with an $AR(1)$-$GARCH(1,1)$ model applied to monthly frequency Treasury Bill rates \cite{Talavera-Tsapin-Zholud-2012-ES}. As for the Chinese market, $AR(1)$-$GARCH(1,1)$ from Ref.~\cite{Engle-1982-Em} is applied to the GDP data in China and the conditional variance is derived to measure the uncertainty \cite{Wang-Song-2014-cnERJ}. However, GDP data is quarterly statistics in China and the monthly data is required in our study. Therefore, the Chinese Macroeconomic Prosperity Indices are employed instead. It contains four indices, which are published by China's National Bureau of Statistics (www.cemac.org.cn). Specially, coincident index reflects the trend of the current economy, involving the industry production, employment, demand and income; leading index is composed of a group of leading indicators and used to predict future economic trends; lagging index is constructed by a group of lagging indicators and it is mostly used to provide the confirmation of the economic cycles; warning index is to divide the macroeconomic state into five levels: ``red'' indicates greatly overheated economy and ``yellow'' indicates overheated economy, while ``green'' suggests the normal economy, ``light blue'' indicates the cold economy, and ``blue'' indicates the much colder economy.
\begin{equation}
\centering
\begin{aligned}
& y_{t}=c+y_{t-1}+\varepsilon_{t},\\
& \varepsilon_{t}=\sigma_{t}z_{t},\\
& \sigma^{2}_{t}=k+\gamma\sigma^{2}_{t-1}+\alpha\varepsilon^{2}_{t-1}.\\
\end{aligned}
\label{Eq:AR-GARCH}
\end{equation}

In our work, the conditional variance associated with an $AR(1)$-$GARCH(1,1)$ model
\cite{Wang-Xie-2016-ESA}
applied to monthly log return of coincident index is used to define the macroeconomic uncertainty, which can be described by Eq.~(\ref{Eq:AR-GARCH}), where $y_{t}$ denotes the log return of coincident index at month $t$.
As with Eq.~(\ref{Eq:GJR}), the conditional mean and variance equations are described by the first and third equations in Eq.~(\ref{Eq:AR-GARCH}), respectively. $c$ and $\varepsilon_{t}$ are the constant and innovation terms, respectively. $\sigma_{t}$ represents the conditional standard deviation of $\varepsilon_{t}$. In addition, $k, \gamma, \alpha>0$ and $\gamma+\alpha<1$.
The high (low) macroeconomic uncertainty regime is defined as periods in which macroeconomic uncertainty is above (below) its median value, and the associated dummy variable equals to unity (zero) in the high (low) macroeconomic uncertainty regime. It is noticeable that the results are robust to the adoptions of other three alternatives, and we merely present the results based on the coincident index.

Table \ref{TB:MarketCondition:CON} reports the performances of CSCON portfolios during the period with both high and low level of each market condition. Apparently, when $J \geq 36$, 6 in 11 CSCON portfolios have statistically significant positive return, thus indicating the long-term CSCON effect in the Chinese stock market \cite{Shi-Jiang-Zhou-2015-PLoS1}.

\setlength\tabcolsep{2pt}
\begin{table}[!ht]
\footnotesize
\caption{
{This table reports the performance of cross-sectional contrarian (CSCON) portfolios based on Fama-French three-factor model with inclusion of market condition dummy, namely, $ER_{t}=\alpha+\beta_{\rm{MKT}}R_{{\rm{MKT}},t}+\beta_{\rm{SMB}}R_{{\rm{SMB}},t}+\beta_{\rm{HML}}R_{{\rm{HML}},t}+\beta_{D}D_{t}+\varepsilon_{t}$ . The dummy variable $D_{t}$ for each market condition takes the value of unity or zero according to the description in the paper. The estimation and holding period is $J$-month and $K$ months, respectively. The values of $J$ and $K$ for different portfolios are indicated in the first two collumns. Monthly average returns of the CSCON portfolios during the whole sample period from 1990 to 2014 are also reported in third column, Raw~Return, against which the risk-adjusted CSCON profitability can compare. The sample period is December 1990 to  December 2014.
Panel A to D reports the results based on four different market conditions that consist of market state (State), market volatility (Volatility), market illiquidity (Illiquidity) and macroeconomics uncertainty (Uncertainty). Following Ref.~\cite{Newey-West-1987-Em}, $t$-statistics are adjusted for heteroscedasticity and autocorrelation. The superscripts * and ** denote the significance at 5\% and 1\% levels, respectively.}}
   \label{TB:ThreeFactorD:CON}
   \centering
   \begin{tabular}{ccccccccccccccc}
   \hline
     $J$ && $K$ && Raw~Return && $\alpha$ && $\beta_{\rm{MKT}}$ &&  $\beta_{\rm{SMB}}$ && $\beta_{\rm{HML}}$ && $\beta_{D}$ \\
   \hline
   \multicolumn{15}{l}{\textit{Panel A : State}} \\
   1 && 1&& {0.0084$^{~~}$}  &&{0.0045$^{~~}$} && {0.0612$^{~~}$} && {\bf 0.2329$^{*~}$} && {-0.0069$^{~~}$} && {0.0005$^{~~}$}  \\
   6 && 1&& {0.0009$^{~~}$}  &&{\bf -0.0132$^{**}$} && {0.0426$^{~~}$} && {\bf 0.2767$^{*~}$} && {0.1253$^{~~}$} && {\bf 0.0189$^{**}$}  \\
   12 && 1&& {0.0044$^{~~}$}  &&{-0.0114$^{~~}$} && {0.0313$^{~~}$} && {0.3109$^{~~}$} && {\bf 0.4965$^{**}$} && {\bf 0.0176$^{*~}$}  \\
   18 && 1&& {0.0057$^{~~}$}  &&{-0.0082$^{~~}$} && {0.0383$^{~~}$} && {\bf 0.4233$^{*~}$} && {\bf 0.6488$^{**}$} && {\bf 0.0174$^{*~}$}  \\
   24 && 1&& {0.0089$^{~~}$}  &&{-0.0057$^{~~}$} && {0.1121$^{~~}$} && {\bf 0.6956$^{**}$} && {\bf 0.5083$^{*~}$} && {0.0133$^{~~}$}  \\
   30 && 1&& {0.0150$^{~~}$}  &&{0.0013$^{~~}$} && {\bf 0.2435$^{*~}$} && {\bf 1.0155$^{**}$} && {0.4150$^{~~}$} && {0.0059$^{~~}$}  \\
   36 && 1&& {\bf 0.0204$^{**}$}  &&{0.0045$^{~~}$} && {\bf 0.2534$^{*~}$} && {\bf 1.0694$^{**}$} && {0.4351$^{~~}$} && {0.0026$^{~~}$}  \\
   42 && 1&& {\bf 0.0228$^{**}$}  &&{0.0035$^{~~}$} && {0.1255$^{~~}$} && {\bf 1.1244$^{**}$} && {0.4488$^{~~}$} && {0.0098$^{~~}$}  \\
   48 && 1&& {\bf 0.0134$^{*~}$}  &&{-0.0085$^{~~}$} && {0.0091$^{~~}$} && {\bf 0.8363$^{**}$} && {\bf 0.5721$^{*~}$} && {\bf 0.0248$^{**}$}  \\
   54 && 1&& {\bf 0.0151$^{**}$}  &&{-0.0085$^{~~}$} && {-0.0014$^{~~}$} && {\bf 0.7878$^{**}$} && {0.3906$^{~~}$} && {\bf 0.0297$^{**}$}  \\
   60 && 1&& {\bf 0.0146$^{*~}$}  &&{-0.0087$^{~~}$} && {-0.0080$^{~~}$} && {\bf 0.8047$^{**}$} && {0.4825$^{~~}$} && {\bf 0.0285$^{**}$}  \\
   \hline
   \multicolumn{15}{l}{\textit{Panel B : Volatility}} \\
   1 && 1&& {0.0084$^{~~}$}  &&{0.0052$^{~~}$} && {0.1041$^{~~}$} && {0.0737$^{~~}$} && {0.0543$^{~~}$} && {0.0039$^{~~}$}  \\
   6 && 1&& {0.0009$^{~~}$}  &&{-0.0057$^{~~}$} && {0.0708$^{~~}$} && {0.1513$^{~~}$} && {0.2140$^{~~}$} && {0.0111$^{~~}$}  \\
   12 && 1&& {0.0044$^{~~}$}  &&{-0.0062$^{~~}$} && {0.1170$^{~~}$} && {0.0643$^{~~}$} && {\bf 0.5937$^{**}$} && {0.0238$^{~~}$}  \\
   18 && 1&& {0.0057$^{~~}$}  &&{-0.0056$^{~~}$} && {-0.0651$^{~~}$} && {0.3105$^{~~}$} && {\bf 0.3994$^{*~}$} && {\bf 0.0209$^{*~}$}  \\
   24 && 1&& {0.0089$^{~~}$}  &&{-0.0039$^{~~}$} && {\bf 0.1662$^{*~}$} && {\bf 0.5297$^{*~}$} && {\bf 0.5214$^{**}$} && {0.0134$^{~~}$}  \\
   30 && 1&& {0.0150$^{~~}$}  &&{-0.0010$^{~~}$} && {\bf 0.2584$^{**}$} && {\bf 0.8287$^{**}$} && {0.3291$^{~~}$} && {0.0095$^{~~}$}  \\
   36 && 1&& {\bf 0.0204$^{**}$}  &&{-0.0013$^{~~}$} && {\bf 0.2472$^{*~}$} && {\bf 1.0597$^{**}$} && {0.4368$^{~~}$} && {0.0182$^{~~}$}  \\
   42 && 1&& {\bf 0.0228$^{**}$}  &&{-0.0018$^{~~}$} && {0.1084$^{~~}$} && {\bf 1.1186$^{**}$} && {0.4438$^{~~}$} && {\bf 0.0283$^{**}$}  \\
   48 && 1&& {\bf 0.0134$^{*~}$}  &&{-0.0030$^{~~}$} && {-0.0077$^{~~}$} && {\bf 0.8501$^{**}$} && {\bf 0.5694$^{*~}$} && {\bf 0.0226$^{*~}$}  \\
   54 && 1&& {\bf 0.0151$^{**}$}  &&{-0.0005$^{~~}$} && {-0.0209$^{~~}$} && {\bf 0.8031$^{**}$} && {0.3910$^{~~}$} && {\bf 0.0236$^{*~}$}  \\
   60 && 1&& {\bf 0.0146$^{*~}$}  &&{0.0006$^{~~}$} && {-0.0242$^{~~}$} && {\bf 0.8190$^{**}$} && {0.4979$^{~~}$} && {0.0195$^{~~}$}  \\
   \hline
   \multicolumn{15}{l}{\textit{Panel C : Illiquidity}} \\
   1 && 1&& {0.0084$^{~~}$}  &&{0.0045$^{~~}$} && {0.1112$^{~~}$} && {0.0763$^{~~}$} && {0.0559$^{~~}$} && {0.0051$^{~~}$}  \\
   6 && 1&& {0.0009$^{~~}$}  &&{-0.0016$^{~~}$} && {0.0748$^{~~}$} && {0.1563$^{~~}$} && {0.2092$^{~~}$} && {0.0017$^{~~}$}  \\
   12 && 1&& {0.0044$^{~~}$}  &&{0.0045$^{~~}$} && {0.1207$^{~~}$} && {0.0743$^{~~}$} && {\bf 0.5807$^{*~}$} && {-0.0001$^{~~}$}  \\
   18 && 1&& {0.0057$^{~~}$}  &&{0.0093$^{~~}$} && {-0.0767$^{~~}$} && {0.3171$^{~~}$} && {\bf 0.3796$^{*~}$} && {-0.0118$^{~~}$}  \\
   24 && 1&& {0.0089$^{~~}$}  &&{0.0042$^{~~}$} && {\bf 0.1632$^{*~}$} && {\bf 0.5368$^{*~}$} && {\bf 0.5111$^{**}$} && {-0.0050$^{~~}$}  \\
   30 && 1&& {0.0150$^{~~}$}  &&{0.0019$^{~~}$} && {\bf 0.2629$^{**}$} && {\bf 0.8365$^{**}$} && {0.3283$^{~~}$} && {0.0020$^{~~}$}  \\
   36 && 1&& {\bf 0.0204$^{**}$}  &&{0.0015$^{~~}$} && {\bf 0.2639$^{*~}$} && {\bf 1.0801$^{**}$} && {0.4485$^{~~}$} && {0.0096$^{~~}$}  \\
   42 && 1&& {\bf 0.0228$^{**}$}  &&{0.0030$^{~~}$} && {0.1376$^{~~}$} && {\bf 1.1471$^{**}$} && {0.4664$^{~~}$} && {0.0134$^{~~}$}  \\
   48 && 1&& {\bf 0.0134$^{*~}$}  &&{0.0054$^{~~}$} && {0.0097$^{~~}$} && {\bf 0.8622$^{**}$} && {\bf 0.5596$^{*~}$} && {-0.0000$^{~~}$}  \\
   54 && 1&& {\bf 0.0151$^{**}$}  &&{0.0047$^{~~}$} && {0.0087$^{~~}$} && {\bf 0.8352$^{**}$} && {0.3759$^{~~}$} && {0.0072$^{~~}$}  \\
   60 && 1&& {\bf 0.0146$^{*~}$}  &&{0.0053$^{~~}$} && {-0.0015$^{~~}$} && {\bf 0.8421$^{**}$} && {0.4785$^{~~}$} && {0.0045$^{~~}$}  \\
   \hline
   \multicolumn{15}{l}{\textit{Panel D : Uncertainty}} \\
   1 && 1&& {0.0084$^{~~}$}  &&{0.0090$^{~~}$} && {0.1030$^{~~}$} && {0.0795$^{~~}$} && {0.0505$^{~~}$} && {-0.0040$^{~~}$}  \\
   6 && 1&& {0.0009$^{~~}$}  &&{0.0019$^{~~}$} && {0.0702$^{~~}$} && {0.1616$^{~~}$} && {0.2057$^{~~}$} && {-0.0054$^{~~}$}  \\
   12 && 1&& {0.0044$^{~~}$}  &&{0.0021$^{~~}$} && {0.1229$^{~~}$} && {0.0695$^{~~}$} && {\bf 0.5827$^{**}$} && {0.0046$^{~~}$}  \\
   18 && 1&& {0.0057$^{~~}$}  &&{0.0050$^{~~}$} && {-0.0629$^{~~}$} && {0.3218$^{~~}$} && {\bf 0.3870$^{*~}$} && {-0.0025$^{~~}$}  \\
   24 && 1&& {0.0089$^{~~}$}  &&{0.0051$^{~~}$} && {\bf 0.1667$^{*~}$} && {\bf 0.5472$^{**}$} && {\bf 0.5144$^{**}$} && {-0.0068$^{~~}$}  \\
   30 && 1&& {0.0150$^{~~}$}  &&{0.0057$^{~~}$} && {\bf 0.2584$^{**}$} && {\bf 0.8429$^{**}$} && {0.3245$^{~~}$} && {-0.0059$^{~~}$}  \\
   36 && 1&& {\bf 0.0204$^{**}$}  &&{0.0059$^{~~}$} && {\bf 0.2528$^{*~}$} && {\bf 1.0717$^{**}$} && {0.4346$^{~~}$} && {0.0002$^{~~}$}  \\
   42 && 1&& {\bf 0.0228$^{**}$}  &&{0.0099$^{~~}$} && {0.1240$^{~~}$} && {\bf 1.1358$^{**}$} && {0.4492$^{~~}$} && {-0.0017$^{~~}$}  \\
   48 && 1&& {\bf 0.0134$^{*~}$}  &&{0.0082$^{~~}$} && {0.0093$^{~~}$} && {\bf 0.8715$^{**}$} && {\bf 0.5582$^{*~}$} && {-0.0062$^{~~}$}  \\
   54 && 1&& {\bf 0.0151$^{**}$}  &&{0.0090$^{~~}$} && {-0.0008$^{~~}$} && {\bf 0.8217$^{**}$} && {0.3774$^{~~}$} && {-0.0025$^{~~}$}  \\
   60 && 1&& {\bf 0.0146$^{*~}$}  &&{0.0073$^{~~}$} && {-0.0071$^{~~}$} && {\bf 0.8317$^{**}$} && {0.4785$^{~~}$} && {-0.0001$^{~~}$}  \\
   \hline
   \end{tabular}
\end{table}

\clearpage

Stark differences for the CSCON profitability occur during upward and downward market state. When the market is in upward trend, 9/11 CSCON portfolios have positive average returns with statistical significance at the level of 5\% and all of them achieve higher profitability, even almost twice their respective raw return. By comparison, the performance of the CSCON portfolios is much poorer when the market is in downward trend. Only CON$(6,1)$ has statistically significant but negative average return.
The results based on market volatility are similar. Likewise, the profitability of CSCON portfolios are more favorable during the period with high level of market volatility. However, only half of them have statistically significant positive average returns when $J \geq 36$. In comparison, only CON$(1,1)$ has statistically significant positive average return during the period with low level of market volatility.
The findings based on market illiquidity and macroeconomic uncertainty are opposite.
As shown obviously in the table, the periods with low market illiquidity and macroeconomic uncertainty are related to higher profitability of CSCON portfolios and they are statistically significant when $J \geq 24$.

The findings above suggest the close relation between the CSCON profitability and market condition factors, which may be owing to the performance of common risk factors. In the following, the augmented FFTM with inclusion of market condition dummy variable is employed to capture the more precise relation between the CSCON profitability and market conditions. The results are presented in Table \ref{TB:ThreeFactorD:CON}.

As for the results based on market state in panel A, none of risk-adjusted average returns exhibit the statistical significance. And the risk-adjusted performances of all CSCON portfolios are worse than their respective raw return. All of loadings on market state dummies are positive, indicating the more superior performance of CSCON portfolios during the period with upward trend of market state. And 6/11 loadings are with statistical significance when $6 \leq J \leq 18$ and $48 \leq J \leq 60$. The results reveal that the CSCON profitability exhibits the dependence on market state, however not to the extent of level implied by the results in Table \ref{TB:MarketCondition:CON}.
Panel B reports the similar results based on market volatility, all of risk-adjusted average returns are insignificant and poorer than their raw returns. Likewise, all of loadings on market volatility dummies are positive, which implies that CSCON portfolios would have more favorable performance during the periods with higher market volatility. However, 4/11 CSCON portfolios are with statistical significance when $J=18$ and $ 42 \leq J \leq 54$. Apparently, the performance of common risk factor would increase the dependence of CSCON profitability on market state and volatility.

Panel C and D report the results based on market illiquidity and macroeconomic uncertainty, respectively.
9/11 loadings on macroeconomics uncertainty dummies are negative but insignificant,
and 4/11 loadings on market illiquidity dummies are negative but insignificant.
Apparently, the findings are not in accordance with those from Table \ref{TB:MarketCondition:CON}. After eliminating the influence of common risk factors, the dependence of CSCON profitability on illiquidity and uncertainty is largely weakened.

According to the results from studies based on two different perspectives, CSCON profitability are highly dependent on the market conditions. The periods with upward trend of market state, higher market volatility, higher market liquidity and lower macroeconomics uncertainty are related to higher CSCON profitability. After allowing for the influence of common risk factors, the relation between CSCON profitability and market conditions are weakened. In summary, the findings are in accordance with the AMH.

\section{Conclusion}
\label{S1:Conclusion}

In this paper, we focus on testing some implications from the Adaptive Markets Hypothesis of Ref.~\cite{Lo-2004-JPM,Lo-2005-JIC} mainly in the Chinese stock markets. We conduct the study to test the practical implications of the AMH within the framework of CSMOM and CSCON effects. We test if the risk-premium relation is stable through checking the performance of CSCON portfolios after risk adjustment by the CAPM and the Fama-French three-factor model. Our findings suggest that the risk-adjusted returns are time-varying, and the loadings on risk factors are not constant, which are consistent with the AMH.

We then conduct a study on the performance of CSCON portfolios in moving windows to verify if the arbitrage opportunities are evolving in Chinese market and other major stock exchanges as well. It is observed that the CSMOM and CSCON effects are evolving over time, as is ubiquitous in all markets under consideration. In addition, we argue that market anomalies characterize the market only in certain periods, which throws light on the some mixed results in related studies.

Finally, we explore the relation between CSCON profitability and market conditions. We evaluate the performance of CSCON portfolios during the period with low and high level of several market conditions, and also investigate if the performance of CSCON portfolios differs from various market environments via augmented Fama-French three-factor model with inclusion of market condition dummy. We find that the CSCON profitability are closely linked with market conditions. The periods with upward trend of market state, higher market volatility, higher market liquidity and lower macroeconomics uncertainty are related to higher CSCON profitability. After allowing for the influence of risk factors, the relation between CSCON profitability and market conditions are weakened. In summary, our findings are in accordance with the AMH.

\section*{Acknowledgements}

This work was partly supported by the National Natural Science Foundation of China (Grant 71532009) and the Fundamental Research Funds for the Central Universities  (222201718006).

\bibliographystyle{elsarticle-num}
\bibliography{E:/papers/Auxiliary/Bibliography}

\end{document}